\documentclass[twocolumn,aps,prb,floatfix,superscriptaddress]{revtex4-2}
\usepackage{psfrag,epsfig,amsfonts,amssymb,amsmath,wasysym,bm}
\usepackage{dcolumn}
\usepackage{bbold}
\usepackage[normalem]{ulem}
\usepackage{color}
\usepackage{tabularx}

\usepackage{enumitem} 

\usepackage{hyperref}

\newcommand{\ket}[1]{\lvert #1 \rangle} 	
\newcommand{\bra}[1]{\langle #1 \rvert}	

\newcommand{\<}{\left\langle}	
\renewcommand{\>}{\right\rangle}	

\newcommand{\pmax}{p_{\mathrm{max}}}
\newcommand{\Ik}{I_{k}}

\newcommand{\Imc}{{\cal I}_{\mathrm{mc}}}

\newcommand{\RR}{{\mathbb R}}
\newcommand{\CC}{{\mathbb C}}
\newcommand{\NN}{{\mathbb N}}

\newcommand{\pu}{\tr\{\rho^2\}}
\newcommand{\prob}{{\cal P}}
\newcommand{\Pg}{P_{\! g}}
\newcommand{\Pc}{\rho_{\mathrm{th}}}

\newcommand{\tr}{\mbox{Tr}}

\newcommand{\hr}{{\cal H}}

\newcommand{\Amc}{\langle A\rangle_{\!\mathrm{mc}}}
\newcommand{\Omc}{\langle O\rangle_{\!\mathrm{mc}}}

\newcommand{\ord}{{\cal O}}
\newcommand{\mic}{\mathrm{mc}}
\newcommand{\rhomic}{\rho_{\mathrm{mc}}}
\newcommand{\rhoth}{\rho_{\mathrm{th}}}

\providecommand{\norm}[1]{\|#1\|}

\newcommand{\lmat}{\left( \begin{matrix}}	
\newcommand{\rmat}{\end{matrix} \right)}	

\newcommand{\DO}{R_{\!\, O}}

\newcommand{\mref}[1]{\ref{#1}}

\definecolor{revcolor}{RGB}{32,64,255}

\begin{document}

\title{Thermalization of locally perturbed many-body quantum systems}
\author{Lennart Dabelow}
\affiliation{RIKEN Center for Emergent Matter Science (CEMS), Wako, Saitama 351-0198, Japan}
\author{Patrick Vorndamme}
\affiliation{Faculty of Physics, 
Bielefeld University, 
33615 Bielefeld, Germany}
\author{Peter Reimann}
\affiliation{Faculty of Physics, 
Bielefeld University, 
33615 Bielefeld, Germany}
\date{\today}

\begin{abstract}
Deriving 
conditions under which a macroscopic
system thermalizes directly 
from the underlying 
quantum many-body dynamics of its microscopic constituents is a long-standing 
challenge in theoretical physics.
The 
well-known
eigenstate thermalization hypothesis (ETH) is presumed to be a key mechanism,
but has defied rigorous verification for 
generic
systems thus far.
A weaker variant (weak ETH), by contrast, is provably true for a 
large variety of systems,
including even many integrable models,
but 
its implications 
with respect to the problem of
thermalization are still largely unexplored.
Here we analytically demonstrate that systems satisfying the weak ETH 
exhibit thermalization for two very natural classes of far-from-equilibrium 
initial conditions:
the overwhelming majority of all pure states with a preset non-equilibrium 
expectation value of some given
local 
observable, and the Gibbs states of a Hamiltonian which subsequently 
is subject to a quantum quench in the form of a 
sudden
change of some local system properties.
\end{abstract}

\maketitle


\section{Introduction}

Statistical mechanics for systems at thermal equilibrium 
is a highly developed cornerstone of theoretical physics. 
Its universal and in principle surprisingly simple 
``working recipe'' is to properly choose and evaluate 
one of the textbook canonical ensembles. 
Even though the considered systems may
be out of equilibrium at the beginning,
these concepts are 
meant to apply to all sufficiently late times, i.e., after 
the initial relaxation processes have died out.
While this prediction or postulate of thermalization
is known to be extremely successful in practice, 
a direct derivation from the basic laws of physics 
is widely considered as a very important and still 
not satisfactorily solved problem.

The most common and natural starting point in this
context is to focus on isolated quantum many-body 
systems, with the objective to explain why they
generically exhibit thermalization in the long run, i.e., 
why they are very well described -- after the relaxation 
of the possibly far-from-equilibrium initial state has 
been completed --
by a microcanonical ensemble
(or by an equivalent canonical ensemble), 
as predicted by the textbooks.
Indeed, this has been the main goal in most of the 
pioneering works in this field 
\cite{neu29,deu91,sre96,tas98}.
A key role in this context is played by the so-called 
eigenstate thermalization hypothesis (ETH)
\cite{neu29,deu91,sre94,rig08}.
Specifically, the 
best-known
or ``strong'' version of the
ETH (sETH) 
can be readily shown to guarantee thermalization 
under rather mild and physically reasonable preconditions 
on the considered observables and initial states
\cite{tas16,ued20,dal16,gog16,mor18}.
The sETH itself, however, still amounts to an 
unproven hypothesis,
and is actually known to be violated by integrable,
many-body localized, and even certain non-integrable 
models \cite{tas16,ued20,dal16,gog16,mor18}.
Accordingly, to analytically deduce thermalization
directly from the unitary quantum dynamics 
remains one of the main challenges
in this research area.

On the other hand, a weaker 
version of the ETH (wETH) has recently been analytically 
established for a large variety of systems
\cite{bir10,mor16,iyo17,mor18,kuw20,kuw20a},
but its implications 
with respect to the issue of thermalization are still not 
very well understood.
This
is the main objective of our present paper:
Focusing on 
cases where 
the wETH is known to apply, we will identify 
two very large and natural classes of non-equilibrium 
initial states for which thermalization can be 
analytically verified,
namely typical pure states with tunable (non-equilibrium) expectation values of local observables in Sec.~\ref{sec:TypicalPureStates}
and local quenches from thermal equilibrium (Gibbs) states in Sec.~\ref{sec:LocalQuenches}.
Notably, while the sETH mainly concerns so-called
non-integrable 
systems, the wETH and 
thus our present results also pertain to 
many integrable models.

\section{Preliminaries}
\label{sec:Preliminaries}

\subsection{Setup}
\label{sec:Preliminaries:Setup}

As announced, 
we consider isolated quantum system with 
$L\gg 1$ degrees of freedom, which are 
moreover known to satisfy the wETH.
We thus focus on translationally invariant 
spin models
with short range interactions,
periodic boundary conditions, 
and for simplicity we restrict ourselves 
to one-dimensional lattices with 
sites $i= 1,...,L$
(various generalizations are straightforward,
see also
Sec.~\ref{sec:Discussion} below).
Accordingly, the Hamiltonian is of the form 
\begin{eqnarray}
H=\sum_{i=1}^L h_i \ ,
\label{zz1}
\end{eqnarray}
where the 
$h_i$ are translational copies of 
the same 
local (few-body and short-range) 
operator, meaning that every $h_i$
only acts nontrivially on lattice sites 
sufficiently close to $i$.
The eigenvalues of $H$ are denoted by $E_n$,
the eigenvectors by $|n\rangle$,
and the underlying Hilbert space by $\hr$, where
$n=1,...,N$ and $N$ is exponentially 
large in the system size $L$.
Furthermore, we 
mainly have in mind observables $O$
which are local operators as specified above
(and are thus sometimes written in the form $O_i$), 
or suitable sums thereof,
as exemplified by the energy (\ref{zz1}).
Focusing on such observables is very common and generally 
considered to still cover most situations of actual interest 
\cite{tas16,ued20,dal16,gog16,mor18}.

As detailed in \cite{bir10,mor16,iyo17,mor18,kuw20,kuw20a},
these Hamiltonians and observables can be shown to satisfy the wETH,
meaning that the empirical variance of the diagonal 
matrix elements $O_{nn}:=\bra{n} O \ket{n}$ taken over macroscopically 
small energy intervals vanishes in the thermodynamic limit
$L\to\infty$ 
[see also Eq.~\eqref{zz3} below].

\subsection{Typical equilibrium states, clustering, and weak ETH}
\label{sec:Preliminaries:TypClustWETH}

Our first objective is to demonstrate thermalization
for a large class of non-equilibrium initial states.
Similarly as in the well-known previous explorations 
of (canonical) typicality and concentration of 
measure phenomena \cite{llo88,gol06,pop06},
we therefore focus on a microcanonical
energy window $\Imc:=[E-\Delta,E]$,
and we denote by $S$ the concomitant 
subset of indices $n$ with  $E_n\in \Imc$,
by $P:=\sum_{n\in S}|n\rangle\langle n|$
the projector onto the so-called energy 
shell (sub-Hilbert space) $\hat \hr$, and 
by $|S|:=\tr\{P\}$ its dimension.
Furthermore, the energy interval $\Imc$ 
can and will be chosen large on microscopic 
and small on macroscopic scales, 
i.e., $|S|$ is still exponential in the system size
$L$, while any (normalized) state 
$|\psi\rangle\in\hat \hr$ exhibits a macroscopically 
well-defined energy (small energy spread).
According to textbook statistical mechanics,
the expectation value of an observable $O$
at thermal equilibrium then follows as 
$\tr\{\rhomic O\}$, where $\rhomic:=P/|S|$ is
the microcanonical ensemble.
Observing that all normalized 
$|\psi\rangle\in\hat \hr$ are of the form 
$\sum_{n\in S} c_n|n\rangle$
with $c_n\in\CC$ and $\sum_{n\in S} |c_n|^2=1$, 
we may view them as points on the unit 
sphere in $\CC^{|S|}$ 
(or
$\RR^{2|S|}$).
If one samples states uniformly from that sphere,
it has been shown for instance in Refs.~\cite{llo88,gol06,pop06}
that they typically amount to equilibrium states 
in the sense that the expectation values 
$\langle O\rangle_{\!\psi}:=\langle \psi|O|\psi\rangle$ 
are very close to the thermal equilibrium value
$\Omc:=\tr\{\rhomic O\}$
for the overwhelming majority of all those $|\psi\rangle$.
More precisely speaking, the difference 
$\langle O\rangle_{\!\psi}-\Omc$
is exponentially small in 
$L$ apart from an exponentially small fraction of 
exceptional $|\psi\rangle$'s.

Importantly, the operator $O$ in the above
typicality result may actually be chosen largely arbitrarily
(it may be non-local and even non-Hermitian).
One thus can conclude that, for instance, also
the correlations 
\begin{eqnarray}
C_{\!\psi}(O_i,O_j'):= \langle O_iO_j'\rangle_{\!\psi} - \langle O_i\rangle_{\!\psi} \langle O_j'\rangle_{\!\psi}
\label{zz2}
\end{eqnarray}
between two local observables $O_i$ and $O'_j$ will be 
exponentially close to
$C_{\!\mic}(O_i,O_j'):=\langle O_iO_j'\rangle_{\!\mic}- \langle O_i\rangle_{\!\mic}\langle O_j'\rangle_{\!\mic}$
for the 
vast majority of all $|\psi\rangle$'s.

Our first remark is that the general mindset of such a 
typicality approach is very natural
from the common information-theoretic 
viewpoint in statistical physics:
Since the actual system state in a real experiment
is usually not exactly known (and in fact not even reproducible),
it is sensible to randomly
sample states $|\psi\rangle$ which conform with the available 
information (here the pertinent energy interval $\Imc$) 
but are otherwise as unbiased as possible.
The above typicality result guarantees that practically 
all those random states then indeed behave 
almost identically (as expected in the real experiment).
The generalization when additional information about 
the initial state is available will be addressed in the 
next section.

Our second remark is that, according to common (textbook) 
wisdom, the standard thermal equilibrium ensembles 
do not
exhibit any unphysical properties.
It follows that the same must apply to the overwhelming
majority of the $|\psi\rangle$'s since they behave practically
indistingushable from a microcanonical ensemble.

Nevertheless,
it might {\em a priori} not
be 
immediately obvious, for example, whether
our simple-minded random states $|\psi\rangle$ 
satisfy the so-called cluster decomposition 
property (CDP), 
which requires
that the correlations (\ref{zz2}) 
must decay to zero with increasingly large distance 
between the lattice sites $i$ and $j$ 
\cite{rig08,mul15,ess16,far17,doy17,sot14,mur19,glu19}.
Indeed, following Weinberg  \cite{wei97}, the CDP is by 
now a well-established 
premise
which any physically
realistic state is supposed to fulfill (at least outside 
the realm where phase transitions may occur).
The justification is considered as essentially self-evident:
Physically realistic states of macroscopic systems, 
as we consider them here, are not expected to
admit correlations between local properties
with large spatial separation.
To formally verify the CDP itself
for a given (pure or mixed) state is in general a 
quite difficult task.
However,
in our present case we can 
exploit that the CDP has been established 
in \cite{ara69,par82,par95,kli14,fro15} 
for thermal Gibbs states (canonical ensembles),
and that the equivalence of ensembles 
has been shown in \cite{tas18} to apply even for 
non-local observables of the form $O_iO_j'$.
We thus can conclude that the CDP 
is also satisfied by our microcanonical 
ensemble $\rhomic$, and finally, according to the 
argument below (\ref{zz2}), also by the overwhelming majority 
of the pure states $|\psi\rangle$.

Note that the above quoted proofs of the CDP and the equivalence of
ensembles
\cite{ara69,par82,par95,kli14,fro15,tas18}
only apply to short-ranged Hamiltonians.
Likewise, our states $\ket\psi$ are sampled from a subspace (energy shell)
which itself encapsulates substantial  information 
about the underlying Hamiltonian's ``locality'' properties.
By contrast,
when sampling random $|\psi\rangle$'s from a high-dimensional
but otherwise arbitrary subspace of $\hr$,
the majority of them may well violate the CDP in general.

Our last remark is that 
we now have all the necessary 
ingredients at
hand
to formally state
the wETH:
Recalling
that
$O_{nn} := \bra{n} O \ket{n}$,
it essentially requires that
\begin{equation}
\label{zz3}
	\Delta_O^{\,2} := 
	\frac{1}{\lvert S \rvert} \sum_{n \in S} \big( O_{nn} - \Omc \big)^{\! 2}
		\to 0
		\
		\mbox{ for}\ L \to \infty
\end{equation}
if one adopts the usual (physically natural) scaling of the microcanonical energy 
window $\Imc$ with the system size $L$ \cite{bir10,mor16,iyo17,mor18,kuw20,kuw20a}.

\section{Thermalization and clustering for typical non-equilibrium initial states}
\label{sec:TypicalPureStates}

\subsection{Result}
\label{sec:TypicalPureStates:Result}

To arrive at our first main result,
we consider a particularly simple and natural extension of 
the above specified 
typicality approach into the non-equilibrium realm. 
Namely, let us assume that, besides the pertinent energy 
interval $\Imc$, also the (possibly far from equilibrium) 
expectation value of some specific local observable $A$ is 
(approximately) known, e.g., because it has been tuned 
experimentally to prepare the system out of equilibrium.
Accordingly, among all the $\ket\psi$'s from before, 
we only keep a subset of states which comply with 
this extra information that $\langle A\rangle_{\!\psi}$ is 
close to some given non-equilibrium 
value $a$.
The ensemble of all these states thus gives 
rise to a statistical operator $\rho$ which is still 
reminiscent of a microcanonical ensemble,
albeit with the additional constraint 
$\tr\{\rho A\} \simeq a$
(akin to a ``generalized microcanonical 
ensemble'' \cite{cas11}).
In particular, we still expect the CDP to hold,
see also the discussion around Fig.~\ref{fig2} below.
To explicitly construct those random states $|\psi\rangle$
and the concomitant ensemble $\rho$, we adopted
the previously developed formalism from 
Refs.~\cite{bar09,rei20}, see also
Sec.~\ref{sec:TypicalPureStates:Derivation}
for further details.

The first main result of our present paper is that
when considering such $|\psi\rangle$'s as initial 
states, which subsequently evolve in time according to 
$|\psi(t)\rangle=e^{-iHt}|\psi\rangle$,
then the overwhelming majority of them exhibits thermalization.
More precisely,
the
time-dependent expectation values 
$\langle O\rangle_{\!\psi(t)}:=\langle\psi (t)|O|\psi(t)\rangle$
of an observable $O$,
which may but need not be equal to $A$, 
stay close to
the thermal equilibrium value
$\Omc$
for 
nearly all
sufficiently late times $t$.
(In particular, not only long-time averages
but single time points are thus considered.)
As usual \cite{dal16,gog16,mor18}, 
some non-small deviations
$\langle O\rangle_{\!\psi(t)}-\Omc$
may still occur during the initial relaxation process
and also at certain arbitrarily late but very rare times 
$t$ (quantum revivals).
Quantitatively, the
deviations $\langle O\rangle_{\!\psi(t)}-\Omc$ are 
predicted 
to decrease 
as $L^{-1/2}$
for the vast majority of 
all (sufficiently late) times $t$ and initial states 
$|\psi\rangle$,
the exceptional $t$'s and $|\psi\rangle$'s being
exponentially rare in $L$.
The derivation will be given in Sec.~\ref{sec:TypicalPureStates:Derivation} below.

\subsection{Example}

\begin{figure}
\includegraphics[scale=1]{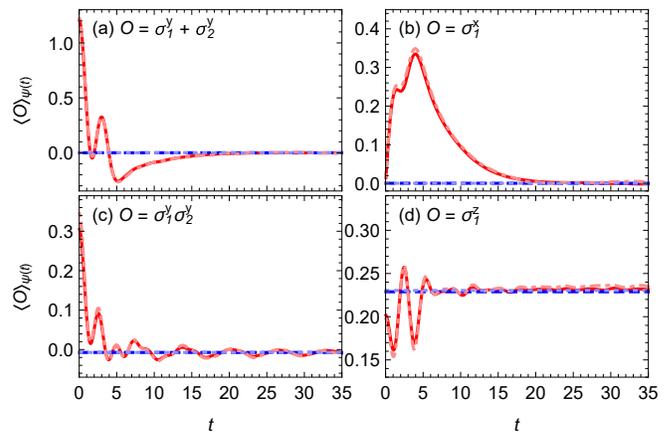}
\caption{Numerical illustration of thermalization
for the integrable model (\ref{zz6}) with 
$L=28$,
$J=\frac{1}{2}$, and $g=\frac{1}{4}\,$.
Red solid: Expectation values $\langle O\rangle_{\!\psi(t)}$ 
of various observables $O$ (see legend;
note the different $y$-axis scales).
Blue dashed: Thermal expectation values of $O$.
Red dash-dotted and blue dotted: Same for $L=24$
(sometimes hardly distinguishable from $L=28$).
As detailed in the main text,
the initial states $|\psi \rangle$ and the thermal expectation 
values were obtained by utilizing a Gaussian filter with parameters 
$\gamma =2.5/L$, $\lambda=-0.35\,L$
(the ground state energy 
is close to $-0.53\, L$),
and by imposing the non-equilibrium initial value
$\langle A\rangle_{\!\psi} \simeq 1.2$
for the observable $A=\sigma^y_1 + \sigma^y_2$ ($=O$ from (a)).}
\label{fig1}
\end{figure}

A numerical illustration of our analytical result 
is provided in Fig.~\ref{fig1} for the common 
transverse-field Ising model (TFIM) 
\cite{dal16,ess16,vid16} with
\begin{eqnarray}
H & = &  -J\, \sum_{i=1}^L \sigma^x_i \sigma^x_{i+1} - g  \sum_{i=1}^L \sigma^z_i
,
\label{zz6}
\end{eqnarray}
where the Pauli matrices $\sigma_i^{\nu}$ with $\nu\in\{x,y,z\}$
and $\sigma^{\nu}_{i\pm L}:=\sigma^{\nu}_i$ (periodic boundary 
conditions) describe the spin components at the chain site $i\in\{1,...,L\}$.
This model obeys the wETH, but violates the sETH,
and is integrable in the sense that
there exists an extensive number of 
local integrals of motion 
(conserved quantities) $\Ik$ 
which commute with $H$ and with each other.
Explicitly, the $\Ik$ with odd
and even $k\geq 1$ are given by \cite{dal16,ess16}
\begin{eqnarray}
I_{2j-1} & = & 
- J\, \sum_{i=1}^L (\tau^{xx}_{i,j+1}+ \tau^{yy}_{i,j-1}) 
+ g\, \sum_{i=1}^L (\tau^{xx}_{i,j}+ \tau^{yy}_{i,j})
,
\nonumber
\\
I_{2j} & = & - J\, \sum_{i=1}^L (\tau^{xy}_{i,j}- \tau^{yx}_{i,j})
\ ,
\label{i1}
\end{eqnarray}
where $\tau^{\nu\mu}_{i,l\geq 1}:=\sigma_i^\nu\sigma_{i+1}^z \cdots \sigma_{i+l-1}^z\sigma_{i+l}^\mu$
and $\tau^{\nu\mu}_{i,0}:=-\sigma_i^z$.

To construct the states $\ket\psi$,
and also 
to determine $\Omc$,
the projector $P$ onto the energy shell
is needed, which
requires the eigenvalues and eigenvectors
of the Hamiltonian $H$.
While these can be explicitly obtained for the present 
system in principle \cite{vid16},
storing enough of them to reach a sensible size of the energy 
shell is beyond our computational resources
for the rather large 
systems
in Fig.~\ref{fig1}.
We therefore followed Ref.~\cite{filter}
(and further references cited therein)
to approximate $P$ by a so-called Gaussian filter
$\Pg:=\exp\{-\gamma (H-\lambda)^2 \}$, 
and analogously for $\Omc$,
see also Appendix~\ref{app:B}
for additional numerical details.
Once a
non-equilibrium
initial state $|\psi\rangle$
has been found,
we numerically determined
its time evolution by means of Suzuki-Trotter product 
expansion techniques \cite{rae04}.
Likewise, $\Pg $ from above was numerically 
obtained via imaginary time evolution.

The 
examples in Fig.~\ref{fig1} nicely confirm 
our general analytical prediction of thermalization.
We also verified (not shown) that the results for different 
randomly sampled initial states $|\psi\rangle$ 
are nearly indistinguishable
(as predicted above Eq.~\eqref{zz4}).
Moreover, the example in Fig.~\ref{fig1} illustrates that,
indeed, thermalization may even occur in integrable systems 
with far-from-equilibrium initial conditions.

\begin{figure}
\includegraphics[scale=1]{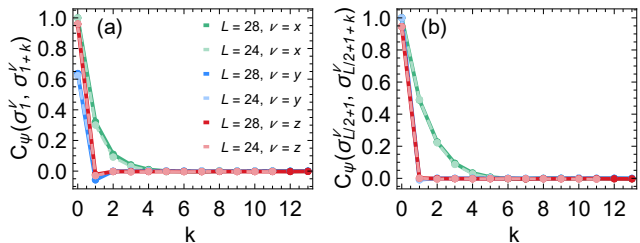}
\caption{Correlation $C_\psi(\sigma^\nu_{j}, \sigma^\nu_{j+k})$ from~\eqref{zz2} 
between local spin operators 
(Pauli matrices $\sigma_i^{\nu}$) as a 
function of their spatial separation $k$,
for the same model Hamiltonian (\ref{zz6}) 
and initial states $\ket\psi$ as in Fig.~\ref{fig1}.
(a) $j=1$, i.e., the first observable $\sigma^\nu_{j}$
appearing in $C_\psi(\sigma^\nu_{j}, \sigma^\nu_{j+k})$
is located at one of the sites which are also affected by the
observable $A=\sigma^y_1 + \sigma^y_2$
that governs the initial non-equilibrium constraint.
(b) $j = L/2+1$, i.e., the first observable $\sigma^\nu_{j}$ appearing 
in $C_\psi(\sigma^\nu_{j}, \sigma^\nu_{j+k})$ 
is located ``far away from $A$''.
}
\label{fig2}
\end{figure}

Finally, one may wonder, similarly as 
in the previous section, whether the states $|\psi\rangle$
still satisfy (with overwhelming probability) the CDP in spite 
of the additional constraint that $\langle A\rangle_{\!\psi}$
must now be close $a$ (see
Sec.~\ref{sec:TypicalPureStates:Result}).
Analogously to the unconstrained case, 
the above mentioned dynamical typicality framework 
 \cite{bar09,rei20} now readily implies 
that this is equivalent to the question whether the
concomitant
ensemble $\rho$ 
exhibits the CDP.
Since analytical results regarding the CDP
are still rather scarce and the few existing proofs 
rather involved, and since this question is not
really a central issue of our present work,
we content ourselves with a numerical illustration, 
shown in Fig.~\ref{fig2} for the same setting as in Fig.~\ref{fig1}:
Both in the vicinity of the ``perturbation'' $A$
(left panel, $j = 1$) and far away from it (right panel, $j = \frac{L}{2}+1$), 
the correlations $C_\psi(\sigma^\nu_{j}, \sigma^\nu_{j+k})$
decay to zero with increasing separation $k$ between the
observables' supports.
Similarly as in Fig. \ref{fig1}, dynamical typicality furthermore 
predicts, and our numerical results (not shown) confirm, that also 
practically any other 
initial state $\ket\psi$ generated in this way exhibits practically 
the same behavior of $C_\psi(\sigma^\nu_{j}, \sigma^\nu_{j+k})$ as 
in Fig.~\ref{fig2}.
Altogether, the numerics thus provides strong evidence that
the CDP is still fulfilled for the vast majority of our
non-equilibrium initial states $|\psi\rangle$.

\subsection{Derivation}
\label{sec:TypicalPureStates:Derivation}

Turning to the derivation of
thermalization for most pure states from the above introduced ensemble of $\ket\psi$'s,
our
starting point is the so-called dynamical typicality 
framework by Bartsch and Gemmer \cite{bar09},
see also Refs.~\cite{rei18a,rei20}.
Concretely, the $\ket\psi$'s are constructed as
\begin{equation}
\label{eq:psi}
	|\psi\rangle:={\cal{N}}\, Q \sum_{n=1}^N z_n \, |n\rangle \,,
\end{equation}
where $|n\rangle$ are the energy eigenstates 
(or any other orthonormal basis of $\hr$)
and $z_n$ are complex numbers,
whose real and imaginary parts are 
given by independent, Gaussian 
distributed random variables of 
zero mean and unit variance.
Moreover,
\begin{equation}
\label{eq:Q}
	Q := P + y P A P
\end{equation}
and ${\cal{N}}$ is a normalization 
constant so that 
$\langle\psi|\psi\rangle =1$.
It follows that $|\psi\rangle\in\hat \hr$, and  
that the ensemble of random states considered in
Sec.~\ref{sec:Preliminaries:TypClustWETH}
is recovered in the special case $y=0$.
Following \cite{bar09}, the purpose of the extra parameter $y$ is  
to account for the additional condition that $\langle A\rangle_{\!\psi}$
must be with high probability close to a fixed value $a$, see
also
Eq.~\eqref{a22x} below.
The statistical operator $\rho$ associated with this ensemble and introduced in Sec.~\ref{sec:TypicalPureStates:Result} is thus
$\rho:=[\,|\psi\rangle\langle\psi |\,]_z$ where the symbol $[\,\cdot\, ]_z$ 
indicates 
the average over all the Gaussian
random numbers $z_n$ from above.

Dynamical typicality \cite{bar09, rei18a, rei20} then
asserts that, for any given $t\geq 0$,
the time-evolved expectation values $\langle O\rangle_{\!\psi(t)}$
for nearly all those $\ket\psi$'s practically coincide with those obtained from 
the time evolution $\rho(t) := e^{-i H t} \rho e^{i H t}$ 
of
$\rho$,
provided that $\rho$ itself is of low purity, i.e.,
$\tr\{\rho^2\} \ll 1$.
As a second ingredient, we 
invoke the well-established fact
\cite{ued20,dal16,gog16,mor18} that the
long-time average of the expectation value
$\tr\{\rho(t) O\}$ can be written
 in the form 
$\tr\{\bar{\rho}\,O\}:=\sum_{n\in S} \rho_{nn} O_{nn}$,
where $\bar\rho$
denotes the so-called
diagonal ensemble
(in case of degeneracies, the eigenstates $|n\rangle$ 
must be chosen so that $O$ is diagonal in the 
corresponding eigenspaces of $H$).
Furthermore, it has been demonstrated, e.g., in
Refs. \cite{equil} that the time dependent expectation values 
$\tr\{\rho(t) O\}$ remain very close to the time-average
$\tr\{\bar{\rho}\,O\}$ for the vast majority of all
sufficiently late times $t$ under quite weak conditions:
Essentially, it is sufficient that the energy differences 
$E_m-E_n$
do not coincide for too many index pairs with $m\not= n$,
which is the case for any generic Hamiltonian $H$
\cite{tas98,sre99,equil},
and that $\tr\{\bar\rho^2\} \ll 1$,
which in turn is guaranteed under 
the same condition $\tr\{\rho^2\} \ll 1$ 
as before \cite{rei20}.
To finally establish thermalization, one has to show that
$\tr\{\bar\rho O\}$ 
practically coincides with
$\Omc$.
We achieve this by rewriting the difference 
$\overline{\Delta O} := \tr\{\bar\rho O\} - \Omc$
as $\sum_{n\in S} \rho_{nn} (O_{nn}-\Omc)$
and then exploiting the Cauchy-Schwarz 
inequality to obtain
\begin{eqnarray}
\overline{\Delta O}^2 & \leq & \Delta_O^{\,2}\,   |S| \sum_{n\in S} (\rho_{nn})^2
\ ,
\label{zz4}
\end{eqnarray}
where $\Delta_O^{\,2}$ is
the wETH characteristic from (\ref{zz3}).
Furthermore, we can upper bound the sum in (\ref{zz4}) by 
$\sum_{m,n=1}^N |\rho_{mn}|^2
=\tr\{\rho^2\}$, yielding
\begin{equation}
\label{zz5}
	\overline{\Delta O}^{\,2} \leq \Delta_O^{\,2}  \, \lvert S \rvert  \, \tr\{\rho^2\} \ .
\end{equation}
Since we focus on systems for which the wETH is fulfilled, 
the quantity $\Delta_O^{\,2}$ approaches zero for large $L$ 
according to (\ref{zz3}).

The remaining task is to establish an $L$-independent upper bound for the quantity $|S|\tr\{\rho^2\}$ appearing 
on the right hand side of (\ref{zz5}) and to show that, 
as a consequence, $\tr\{\rho^2\} \ll 1$
for asymptotically large $L$
as required twice above (\ref{zz4}).

Obviously, the
operator $PAP=:\hat A_r$ in~\eqref{eq:Q} is the projection/restriction 
of the original observable $A$ to the energy shell $\hat \hr$ (hence the ``hat'' symbol).
Possibly after adding a trivial constant to the observable $A$, and then 
multiplying it by a constant factor, we can 
assume without loss of generality that
$\hat A_r$ has been ``rescaled'' (hence the index ``$r$'') so that
$\tr\{\rhomic \hat A_r\}=0$ and $\tr\{\rhomic \hat A_r^2\}=1$ \cite{bar09}.
Finally, the eigenvalues of $\hat A_r$ are denoted 
as $a_n$ and the $k$-th moment 
of the eigenvalue distribution as
\begin{eqnarray}
m_k:=\frac{1}{|S|}\sum_{n\in S} (a_n)^k=\frac{1}{|S|}\tr\{\hat A_r^k\}=\tr\{\rhomic \hat A_r^k\}
\ . \ \
\label{a1}
\end{eqnarray}
The above mentioned rescaling thus implies
$m_1=0$ and $m_2=1$, and
$Q$ from~\eqref{eq:Q} can be rewritten as $P+y\hat A_r$.

One of the main results obtained in \cite{rei20} is 
that $\rho$ can be approximated arbitrarily well by
$\tilde \rho:=Q^2/\tr\{Q^2\}$ provided 
$\tr\{\tilde \rho^2\}$ is sufficiently small.
In the following, we therefore tacitly replace $\tilde\rho$ by $\rho$
and subsequently verify that $\tr\{\rho^2\}\ll 1$.
By means of a straightforward but somewhat tedious
calculation (working in the eigenbasis of $\hat A_r$)
one thus can infer that
\begin{eqnarray}
\rho = \frac{1}{|S|}\frac{1}{1+y^2}(P+y\hat A_r)^2
\ .
\label{a2}
\end{eqnarray}
Moreover, the expectation value of $\hat A_r$ is found to be
\begin{eqnarray}
\tr\{\rho \hat A_r\} = \frac{2y+m_3 y^2}{1+y^2}
\ ,
\label{a3main}
\end{eqnarray}
and for the purity of $\rho$ one obtains
\begin{eqnarray}
\tr\{\rho^2\} = \frac{1}{|S|} \frac{1+6y^2+4m_3y^3+m_4y^4}{(1+y^2)^2}
\ .
\label{a4main}
\end{eqnarray}

Generally speaking, the eigenvalues and eigenvectors
of the restricted and rescaled operator $\hat A_r$ 
have little to do with those of the original observable 
$A$.
Yet it seems reasonable to expect that
the eigenvalue distribution of $\hat A_r$  does not exhibit 
long tails so that, given its first two moments 
are $m_1=0$ and $m_2=1$,
also the next two moments $m_3$ and $m_4$
will be (at most) on the order of unity,
see also \cite{ric20} 
for a numerical example.
We thus can conclude that by choosing suitable 
parameter values $y$,
expectation values in (\ref{a3main}) of up to the order of 
unity (in modulus) can be generated \cite{bar09}.
Furthermore, the purity in (\ref{a4main}) is (for any $y$)
on the order of $1/|S|$, which in turn is
exponentially small in the system size $L$ 
(see
Sec.~\ref{sec:Preliminaries:TypClustWETH}).

Altogether, we can conclude that
$|S|\tr\{\rho^2\}$ is upper bounded by an
$L$-independent constant and that
$\tr\{\rho^2\}$ is exponentially small in $L$,
which completes our demonstration of thermalization.
For the remaining quantitative details
mentioned at the end of Sec.~\ref{sec:TypicalPureStates:Result},
namely the precise
scaling 
of the fraction of exceptional $\ket\psi$'s and  the 
deviations $\< O \>_{\!\psi(t)} - \Omc$,
we refer to Appendix~\ref{app:A}.

We close with two side remarks:
First, one may wonder why only local observables
$A$ are admitted in Ref. \cite{bar09} and in the
above considerations. In fact, one can show that
$A$ may actually also consist of any linear combination 
of local operators, as long as the number of summands
remains small.
On the other hand, if $A$ is, for example, an extensive 
observable,
as exemplified by the energy in 
(\ref{zz1}), then 
notable deviations of 
$\langle A\rangle_{\!\psi}$ from the thermal 
equilibrium value $\Amc$ 
can no longer be achieved
by means of our present general framework
for the following reason:
By exploiting (\ref{a3main}) and the dynamical typicality
formalism from Refs. \cite{rei20}, one can show that
the expectation value $\< A \>_{\!\psi}$
of the original observable $A$
is exponentially likely to be exponentially close to
\begin{eqnarray}
\tr\{\rho A \} = \Amc + s \,  \frac{2y+m_3 y^2}{1+y^2}
\ ,
\label{a22x}
\end{eqnarray}
where $0\leq s\leq\sqrt{\tr\{ \rhomic (A - \Amc)^2 \}}$
(thermal fluctuations of $A$).
As a consequence, the typical relative deviations of  $\< A \>_{\!\psi}$ 
from the thermal equilibrium value $\Amc$ 
must remain very small if $A$ is an extensive observable.

Second,
the above mentioned exponentially likely proximity of $\< A \>_{\!\psi}$ to $\tr\{\rho A\}$ 
does not yet exclude the existence of a very
unlikely subset of $|\psi\rangle$'s with substantial 
deviations of $\langle A\rangle_{\!\psi}$ 
from $\tr\{\rho A\}$.
On the other hand, one readily verifies that this exceptional
subset can be excluded from the total
set of admitted $|\psi\rangle$'s right from the beginning,
entailing only exponentially small corrections in all the 
subsequent calculations.
This extra step has been tacitly taken for granted
at the beginning of the
Sec.~\ref{sec:TypicalPureStates:Result}.

\section{Thermalization of a quenched Gibbs state}
\label{sec:LocalQuenches}

\subsection{Result}

We now
turn to the second main result of our paper:
Let us assume that the initial state
is given by a thermal Gibbs state (canonical ensemble) 
of the form 
\begin{eqnarray}
\rho := Z_0^{-1} e^{-\beta H_0}\ ,\ \ Z_0:=\tr\{ e^{-\beta H_0} \}
\ ,
\label{zz7}
\end{eqnarray}
where $H_0$ is different from the Hamiltonian 
$H$ in (\ref{zz1}) which governs the subsequent temporal 
evolution.
(Note that the restrictions on $H$, such as translational 
invariance, do not apply to $H_0$.)
In other words, 
the system is in thermal equilibrium for 
$t<0$ and is subject to an instantaneous
``quantum quench'' at $t=0$, with pre-quench 
Hamiltonian $H_0$ and post-quench Hamiltonian $H$.
Moreover, we focus on so-called local quenches,
meaning that the difference $V := H - H_0$ between the 
post- and pre-quench Hamiltonians is a local operator.
Note that, despite their apparent ``smallness'' (compared to 
$H_0$ and $H$),
it has been observed numerically in a somewhat different context 
that such local perturbations of $H_0$  can
still have profound effects, for instance changing
a system violating the sETH into one satisfying 
it \cite{tor14,bre20,san20}.
Likewise, such local quenches may still give rise
to far-from-equilibrium initial expectation values
(see Fig.~\ref{fig3}(a)).

From a different standpoint, since the change of the Hamiltonian 
is restricted to a finite subsystem,
the situation immediately after the quench 
may also be viewed as a ``small'' non-equilibrium
system in contact with a ``large'' thermal bath.
Although it may seem intuitively reasonable to expect thermalization 
for such a setup, 
verifying this by analytical means is nevertheless an important and 
challenging problem, whose solution is the second main 
achievement of our present paper. 

Namely, assuming again that $H$ obeys the 
wETH~\eqref{zz3},
we can demonstrate that the time-evolved expectation values 
$\tr\{ \rho(t) O \}$, where $\rho(t) := e^{-i H t} \rho \, e^{i H t}$
with $\rho$ from (\ref{zz7}), 
are practically indistinguishable from $\Omc := \tr\{ \rhomic O \}$ 
for nearly all sufficiently late times $t$,
where $\rhomic$ 
is the microcanonical post-quench ensemble
with appropriate energy $\tr\{\rhomic H\}=\tr\{ \rho H \}$.
This result will be derived in Sec.~\ref{sec:LocalQuenches:Derivation} below.

\subsection{Example}

\begin{figure}
\includegraphics[scale=1]{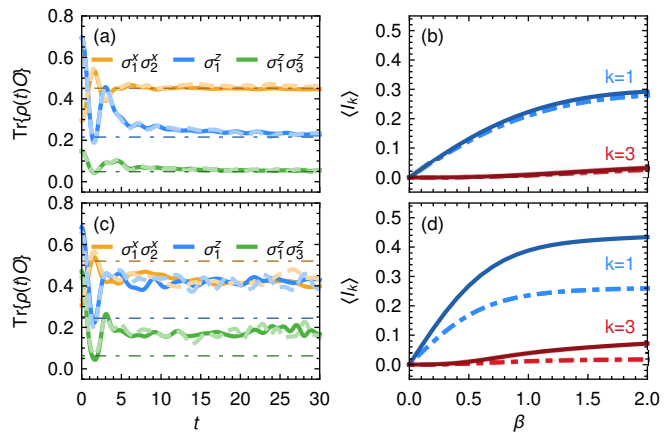}
\caption{Thermalization (top) and non-thermalization (bottom)
after a canonical quench.  
The post-quench Hamiltonian $H$ is given by
the integrable model (\ref{zz6})
with $J=\frac{1}{2}$ and $g=\frac{1}{4}$.
The pre-quench Hamiltonian $H_0$ is given by
$H - \frac{3}{4} (\sigma^z_1 + \sigma^z_2)$
in the top panels 
(local quench such that the total transverse field is $g=1$ for 
the first two sites, but $g = \frac{1}{4}$ for all other sites),
and by $H$ from (\ref{zz6}) with $J=\frac{1}{2}$ and $g=1$ in the bottom panels 
(global quench).
In both cases, a canonical initial state $\rho := e^{-\beta H_0}/\tr\{e^{-\beta H_0}\}$
(see (\ref{zz7}))
was numerically explored by utilizing
dynamical typicality \cite{rei20} and Suzuki-Trotter product 
expansion techniques \cite{rae04},
while
$\rho' := e^{-\beta'\! H}/\tr\{e^{-\beta'\! H}\}$
with $\beta'$ so that $\tr\{ \rho' H \}=\tr\{\rho H \}$
was used as thermal reference ensemble,
see also main text.
(a,c): Time-dependent expectation values 
of three observables (see legend)
for $\beta=1$.
Solid: $L = 24$;
dashed: $L = 20$;
dash-dotted: 
thermal values $\tr\{\rho'O\}$
(indistinguishable for $L=24$ and $L=20$).
(b,d): 
Conserved quantities $I_k$ 
from (\ref{i1})
for 
$k=1,3$
versus $\beta$.
[Graphs for even $k$ are very close to zero and therefore not shown.]
Solid:
conserved expectation values $\tr\{\rho(t)I_k\}=\tr\{\rho I_k\}$.
Dash-dotted:
thermal
values $\tr\{ \rho' I_k \}$.
The system size is $L = 24$ (data for $L = 20$ 
would be indistinguishable on the given scale).}
\label{fig3}
\end{figure}

Our analytical prediction of thermalization after local quenches
is exemplified by Fig.~\ref{fig3}(a,b)
for the TFIM from~\eqref{zz6}.
In particular, these numerical findings illustrate 
that even integrable systems
may thermalize after a local quench.
(For numerical convenience, we actually employed not the 
microcanonical but the equivalent canonical ensemble 
\cite{kuw20,kuw20a,tou15,bra15,tas18,rig14,rig16,mal18}
to evaluate the thermal expectation values,
see also the discussion below Eq.~(\ref{i1})).
Moreover, Fig.~\ref{fig3}(a) shows that
local observables may initially still 
exhibit far-from-equilibrium expectation values,
whereas the conserved quantities in Fig.~\ref{fig3}(b) 
indeed assume the pertinent (time-independent)
thermal values, as it must be.
Remarkably, the results in Fig.~\ref{fig3}(a,b)
are (nearly) $L$-independent, i.e., the large-$L$ 
asymptotics can be anticipated without a sophisticated 
finite size analysis.

For comparison, Fig.~\ref{fig3}(c,d)
also illustrates the effects of a global quench, 
resembling those of a local quench for short times, 
but exhibiting non-thermalization in the long run, 
as expected for the integrable post-quench 
Hamiltonian $H$ at hand (similar examples can also be found in 
Refs.~\cite{rig14,rig16, mal18}).

With regard to related previous works
we remark that
thermalization might also be inferred from
Refs.~\cite{sot14,mur19,glu19}
for the example from Fig.~\ref{fig3}(a,b), 
since this specific post-quench 
Hamiltonian $H$ amounts to a so-called 
non-interacting integrable model.
Furthermore,
as far as $\beta$-values below $0.386$ are concerned, 
thermalization might be understood
by combining the findings from 
Ref.~\cite{far17} with those of Ref.~\cite{kli14}
(and some additional calculations which we omit).
By contrast, our present analytical result is not restricted to 
non-interacting integrable models
or small $\beta$-values.
Finally, the first steps of our derivation
(see below) are also somewhat  reminiscent of 
Theorem~3 in \cite{mul15}, 
Sec.~6 in \cite{mor16}, Sec.~6 in \cite{tas16}, or Theorem~1 in \cite{dun21}, 
but not the more demanding subsequent 
steps
(beginning with Eq.~\eqref{q2}, in particular).

\subsection{Derivation}
\label{sec:LocalQuenches:Derivation}

The first
tasks
in our derivation of
thermalization after local quenches
are 
exactly as in Sec.~\ref{sec:TypicalPureStates:Derivation}, except that 
$\rho$ is now given by (\ref{zz7}):
We have to show that $\tr\{\rho^2\}\ll 1$ and that
$\overline{\Delta O} := \tr\{\bar\rho O\} - \Omc$
is negligibly small.
Recalling that the textbook free energy $F_\beta$
associated with the Gibbs ensemble (\ref{zz7}) 
obeys the relation $e^{-\beta F_\beta}=\tr\{e^{-\beta H_0}\}$,
one can conclude that $\tr\{\rho^2\}=e^{-2\beta \Delta_\beta}$
with $\Delta_\beta:=F_{2\beta}-F_{\beta}$.
Taking for granted that the system exhibits 
generic thermodynamic properties \cite{tas16}, 
it follows that $\Delta_\beta$ is an extensive quantity, 
hence $\tr\{\rho^2\}$ decreases exponentially 
with $L$.
Turning to $\overline{\Delta O}$, one readily sees, 
similarly as in Sec.~\ref{sec:TypicalPureStates:Derivation}, that
\begin{eqnarray}
\overline{\Delta O} & = & X_1+X_2
\ ,
\label{zz8}
\\
X_1 & := &  \sum_{n=1}^N \rho_{nn} (O_{nn}-\Omc')
\ ,
\label{zz9}
\\
X_2 & := & \Omc'-\Omc \ ,
\label{zz10}
\end{eqnarray}
where $\Omc' := \tr\{ \rhomic' O \}$ and $\rhomic'$ is the microcanonical 
ensemble that reproduces the energy of the 
\emph{post-quench} canonical ensemble $\rhoth$ given by
\begin{eqnarray}
	\rhoth &:= & Z^{-1} \, e^{-\beta H} = \sum_{n=1}^N p_n \ket{n} \bra{n} \ ,
	\label{zz11}
	\\
	p_n &:= & Z^{-1} \, e^{-\beta E_n} \ ,
	\label{zz12}
	\\
	Z &:= & \tr\{ e^{-\beta H} \} =  \sum_{n=1}^N e^{-\beta E_n} \ ,
	\label{zz13}
\end{eqnarray}
i.e., $\tr\{\rhomic' H\} = \tr\{ \rhoth H \}$.
Given that $H$ and $H_0$ only differ by a local operator (see above),
it is reasonable to expect that the relative difference between the 
two energies $\tr\{ \rhoth H \}$ and  $\tr\{ \rho H_0 \}$ approaches zero
for large $L$, and likewise for the concomitant difference in (\ref{zz10}).
The quite arduous analytical confirmation of this expectation can be found 
in
Appendix~\ref{app:C}.

The remaining task is to upper bound $X_1$ in (\ref{zz9}).
To this end, we split each summand in (\ref{zz9}) into a factor
$\rho_{nn}/p_n^{1/2}$ and a factor $p_n^{1/2} (O_{nn}-\Omc')$, 
and then invoke the Cauchy-Schwartz inequality to conclude that
\begin{eqnarray}
X_1^2 & \leq & c' \Delta_O^{\prime \, 2}
\ ,
\label{zz14}
\\
c' &:= & \sum_{n=1}^N (\rho_{nn})^2/p_n
\label{zz14a}
\ ,
\label{zz15}
\\
\Delta_O^{\prime \,2}  &:= & \sum_{n=1}^N p_n \big( O_{nn} - \Omc \big)^{\! 2} \ .
\label{zz16}
\end{eqnarray}
One readily recognizes quite considerable similarities to the 
discussion around (\ref{zz4}):
First, $\Delta_O^{\prime\,2}$ obviously amounts to the canonical counterpart
of the corresponding microcanonical quantity 
$\Delta_O^{\,2}$ from (\ref{zz3}).
Second, since we focus on systems for which the wETH is 
fulfilled, we thus can again conclude that $\Delta_O^{\prime\,2}$
approaches zero for large $L$,
see also Appendix~\ref{app:B:Thermalization}
for more details.
Finally, we are again left to show that $c'$ 
(which is obviously the canonical counterpart of
$|S| \sum_{n\in S} (\rho_{nn})^2$ in (\ref{zz4}))
remains bounded for large $L$.

To do so, we
start by considering the operator-valued function 
\begin{eqnarray}
\Phi(\lambda) :=  e^{\lambda H/2} \, e^{-\lambda H_0} \, e^{\lambda H / 2}
\ .
\label{q2}
\end{eqnarray}
One readily verifies that $\Phi(\lambda)$ is Hermitian and that
the derivative $\Phi'(\lambda):=d\Phi(\lambda)/d\lambda$ can be written as
\begin{eqnarray}
\Phi'(\lambda)
& = & 
e^{\lambda H/2} \frac{H-H_0}{2} e^{-\lambda H_0}  e^{\lambda H / 2}
\nonumber
\\
& + &
e^{\lambda H/2}e^{-\lambda H_0}  \frac{H-H_0}{2}  e^{\lambda H / 2}
\ .
\label{q3}
\end{eqnarray}
Exploiting (\ref{q2}), $V:=H-H_0$, and 
\begin{equation}
	V_\lambda := e^{-\lambda H/2} \, V \, e^{\lambda H / 2} \,,
	\label{q4}
\end{equation}
we thus arrive at
\begin{eqnarray}
\Phi'(\lambda)
& = & \frac{V_\lambda^\dagger \Phi(\lambda) + \Phi(\lambda) V_\lambda}{2}
\ .
\label{q5}
\end{eqnarray}
Rewriting (\ref{zz7}) with (\ref{q2}) as 
\begin{eqnarray}
\rho = Z_0^{-1}  e^{-\beta H/2} \, \Phi(\beta) \, e^{-\beta H / 2}
\label{q9}
\end{eqnarray}
and recalling that $|n\rangle$ and $E_n$ are the eigenvectors and
eigenvalues of $H$, it follows that 
\begin{eqnarray}
\langle n|\rho |n \rangle = Z_0^{-1}  e^{-\beta E_n} \, \langle n| \Phi(\beta) | n \rangle 
\ .
\label{q10}
\end{eqnarray}
With (\ref{zz12}) we thus can rewrite (\ref{zz15}) as
\begin{eqnarray} 
c' & = & \sum_{n=1}^N \frac{Z^2}{Z_0^2} \,  p_n  \, \langle n| \Phi(\beta) | n \rangle^2
\nonumber
\\
& \leq & \frac{Z^2}{Z_0^2}  \max_n \langle n| \Phi(\beta) | n \rangle^2  \sum_{n=1}^N  p_n
\nonumber
\\
& = &  \frac{Z^2}{Z_0^2}  \max_n \langle n| \Phi(\beta) | n \rangle^2
\ .
\label{q11}
\end{eqnarray}

To proceed, we utilize that $H$ in (\ref{zz1})
is a sum of local operators $h_i$,
whose operator norms 
$\norm{h_i}$ can be bounded from above 
by an $i$- and $L$-independent constant.
Denoting the support of $h_i$ (lattice sites on which $h_i$ acts nontrivially) 
by $\operatorname{supp}(h_i)$, there must also exist an $L$-independent 
constant $R$ such that 
$\max \{ \lvert x - y \rvert : x, y \in \operatorname{supp}(h_i)\} \leq R$ 
for all $h_i$.

Next we observe that $V_\lambda$ from (\ref{q4})  can be understood 
as an imaginary-time evolution of the observable $V$ with the Hamiltonian $H$.
Since $V$ is a local operator, its support $\operatorname{supp}(V)$ consists 
of a finite number of lattice sites $s_V := \lvert \operatorname{supp}(V) \rvert$.
For real-time evolution,
it is well known that Lieb-Robinson bounds \cite{lie72, has10} limit the growth 
of the support of the time-evolved observable $e^{i H t} V e^{-i H t}$ with $t$ 
(up to exponentially decaying corrections).
As far as our one-dimensional models (\ref{zz1}) are concerned,
similar bounds for the complex-time evolution as in~\eqref{q4}
were first derived by Araki \cite{ara69},
with the difference that the ``light cone'' grows exponentially with the 
magnitude $\lvert \lambda \rvert$ of the complex time
(as opposed to linear growth in Lieb-Robinson-type bounds).
For our purposes, the following formulation based on Bouch \cite{bou15} is
particularly convenient:
For fixed $\lambda$ and any integer $k > e^{c_1 \lvert \lambda \rvert}$,
we can decompose
\begin{equation}
\label{eq:V:iTimeEvo:LocalDecomposition}
	V_\lambda = V_\lambda^{(k)} + \Delta_\lambda^{(k)}
\end{equation}
such that 
$\operatorname{supp}(V_\lambda^{(k)})$ extends at most $k$ sites beyond $\operatorname{supp}(V)$
(meaning that for any $x \in \operatorname{supp}(V_\lambda^{(k)})$ there exists 
a $y \in \operatorname{supp}(V)$ with $\lvert x - y \rvert \leq k$)
and
\begin{equation}
\label{eq:V:iTimeEvo:NonlocalBound}
	\norm{ \Delta_\lambda^{(k)} } \leq s_V \norm{V} \, c_2 \, e^{-k} \, ,
\end{equation}
where $c_1, c_2 > 0$ are $L$-independent constants.
Moreover, for sufficiently large $L > 2(k - s_V)$,
$V_\lambda^{(k)}$ is independent of $L$.
Consequently, $\norm{V_\lambda}$ can be bounded from above by an $L$-independent constant.
Since $V_\lambda$ is furthermore continuous in $\lambda$ (even as $L \to \infty$ \cite{bou15}),
there also exists a common, $L$-independent upper bound $M_\beta$ for all $\lambda \in [0, \beta]$,
i.e.
\begin{equation}
\label{eq:V:iTimeEvo:Bound}
	\sup \{ \norm{V_\lambda} : \lambda \in [0, \beta] \} \leq M_\beta \,.
\end{equation}
In addition, by the same arguments, also $\norm{V_\lambda^\dagger} \leq M_\beta$ 
($\lambda \in [0, \beta]$) for the same constant $M_\beta$.
(Note that we can thus assume $\beta > 0$ in the following without loss of 
generality because $V_{-\lambda} = V_\lambda^\dagger$.
Moreover,
a non-vanishing temperature $\beta^{-1}$ is tacitly taken for granted.)

Observing that $|\norm{B}-\norm{C}|\leq \norm{B-C}$ for arbitrary operators $B$, $C$, 
and choosing $B=\Phi(\lambda +d\lambda)$ and $C=\Phi(\lambda)$ it readily follows that
\begin{eqnarray}
\left| \frac{d\, \norm{\Phi(\lambda)}}{d\lambda}\right| \leq \norm{\Phi'(\lambda)}
\ .
\label{q12}
\end{eqnarray}
With (\ref{q5}) and (\ref{eq:V:iTimeEvo:Bound}) this implies
\begin{eqnarray}
\left| \frac{d\, \norm{\Phi(\lambda)}}{d\lambda}\right| \leq \norm{V_\lambda}
 \norm{\Phi(\lambda)}\leq M_\beta  \norm{\Phi(\lambda)}
\label{q13}
\end{eqnarray}
for all $\lambda\in[0,\beta]$.
Upon integrating this inequality and exploiting
that $\norm{\Phi(0)}=1$ according to (\ref{q2}), we finally obtain
the $L$-independent upper bound
\begin{eqnarray}
\norm{\Phi(\beta)}\leq e^{\beta M_\beta}
\ .
\label{q14}
\end{eqnarray}

Since $|\langle n| \Phi(\beta) | n \rangle|\leq\norm{\Phi(\beta)}$ for all $n$, we
can infer from (\ref{q11}) and (\ref{q14}) that
\begin{eqnarray}
c' &  \leq & \frac{Z^2}{Z_0^2} e^{2 \beta M_\beta}
\ .
\label{q15}
\end{eqnarray}
Evaluating $Z_0$ in (\ref{zz7}) by means of the basis $|n\rangle$, we obtain
\begin{eqnarray}
Z_0=\sum_{n=1}^N \langle n|e^{-\beta H_0}|n\rangle=
\sum_{n=1}^N Z_0 \langle n|\rho|n\rangle
\ .
\label{q16}
\end{eqnarray}
Exploiting (\ref{q10}) on the right hand side of (\ref{q16})
implies
\begin{eqnarray}
Z_0=\sum_{n=1}^N e^{-\beta E_n} \, \langle n| \Phi(\beta) | n \rangle 
\ ,
\label{q17}
\end{eqnarray}
and similarly as in (\ref{q11}), (\ref{q15}) it follows with (\ref{zz13}) that
\begin{eqnarray}
Z_0\leq Z\, e^{\beta M_\beta}
\ .
\label{q18}
\end{eqnarray}
Upon exchanging the roles of $H_0$ and $H$, one analogously finds that
\begin{eqnarray}
Z\leq Z_0\, e^{\beta M_\beta}
\ .
\label{q19}
\end{eqnarray}
With (\ref{q15}) we thus arrive at
the $L$-independent upper bound
\begin{eqnarray}
c' &  \leq & e^{4 \beta M_\beta}
\ .
\label{q20}
\end{eqnarray}

\section{Discussion and conclusions}
\label{sec:Discussion}

In conclusion,
we analytically demonstrated thermalization
for a large variety of non-equilibrium initial conditions with
local perturbation traits, including the overwhelming 
majority of initial states with a preset expectation value of 
some local observable and initial Gibbs states after a local 
quantum quench.
For instance, this may describe a small subsystem
far from equilibrium in contact with a thermal bath
(rest of the system).
Besides focusing -- as usual -- on local observables (and suitable sums thereof),
the (post-quench) Hamiltonian is required to obey the wETH, 
which
has been proven,
among others, for many very common
translationally invariant models with short-range interactions
\cite{bir10,mor16,iyo17,mor18,kuw20,kuw20a}.
In particular, non-integrable as well as integrable models are admitted.
As an example, we numerically illustrated our prediction of 
thermalization for the integrable TFIM (\ref{zz6}), 
and also the previously known absence of 
thermalization after a global quench.

Moreover, these many-body initial states were demonstrated 
to obey the cluster decomposition property, i.e., 
they 
are not marred by
unphysical ``non-locality'' features in the form of
correlations  over large distances.

To keep things simple, we focused on one-dimensional models
and employed the language of spin models,
but fermionic systems or particle-number-conserving 
bosonic systems can be readily transformed into spin models preserving the local 
structure (in one dimension)
and are thus equally covered.
In higher dimensions, generally speaking, additional 
complications like phase transitions will unavoidably
occur at least for some members of the admitted model class.
As a consequence, already the wETH itself,
which we utilized as an ingredient in our present
approach, is only known to be true for energies sufficiently 
far above the realm where such systems in principle may 
exhibit a phase transition 
\cite{bir10,mor16,iyo17,mor18,kuw20,kuw20a}.
Similar restrictions apply to the known proofs of the
cluster decomposition property for thermal Gibbs 
states  \cite{ara69,par82,par95,kli14,fro15}, 
which served as another ingredient of our present explorations.
Nevertheless, provided that those two preconditions are met,
our results regarding thermalization and clustering for typical 
non-equilibrium pure states can be readily adapted as well.
In turn, the imaginary-time analogs of the Lieb-Robinson 
bounds, which we employed in our analysis of the
initial Gibbs states, are until now not available in higher 
dimensions.
Altogether, developing a common framework for
thermalization and phase transitions thus remains
as a very challenging task for future research.

Finally, we point out that the wETH is essential in those 
findings in the sense that systems violating it (e.g., many-body-localized ones) 
can generally not be expected to thermalize from similar initial conditions.

To conclude, the ubiquity of thermalization is a very well established 
empirical observation in numerical and real-world experiments, 
and has the status of an axiom in textbook (equilibrium) statistical 
mechanics.
Our present results may be considered as a notable step forward 
in the long-standing but still largely unsolved task to theoretically 
explain this empirical observation directly from the underlying microscopic 
dynamics.

\begin{acknowledgments}
We thank Stefan Kehrein, J\"urgen Schnack, and Masahito Ueda
for stimulating discussions.
This work was supported by the 
Deutsche Forschungsgemeinschaft (DFG)
within the Research Unit FOR 2692
under Grants No.~355031190
and~397300368,
by the Paderborn Center for Parallel 
Computing (PC$^2$) within the Project 
HPC-PRF-UBI2,
and by the International Centre for Theoretical Sciences 
(ICTS) during a visit for the program -  
Thermalization, Many body localization and Hydrodynamics 
(Code: ICTS/hydrodynamics2019/11).
\end{acknowledgments}


\appendix

\section{Typicality, equilibration, and thermalization}
\label{app:A}

In this appendix,
we provide the quantitative details
regarding thermalization of typical pure states.
We heavily draw
on previously established analytical results 
in the context of ``dynamical typicality''.
In doing so, we mainly employ the most 
general version of 
this formalism from Ref.~\cite{rei20}, which unifies and extends
a considerable number of important precursory works
(see references therein).

Following Ref.~\cite{rei20}, our starting point is
an $N$-dimensional Hilbert space $\hr$, spanned by some 
orthonormal basis
$\{|\chi_n\rangle\}_{n=1}^N$.
For instance, the $|\chi_n\rangle$ may 
be the eigenvectors of some Hamiltonian 
$H$.
In any case (see also main text), 
we focus on many-body systems
with a large but finite number $L$ of degrees
of freedom, and the Hilbert space 
dimension $N$ is understood to be 
exponentially large in $L$ but finite.
Generalizations to infinite $N$
are straightforward \cite{rei20},
but omitted here in order to avoid
inessential technicalities.

\subsection{Dynamical typicality}
\label{app111}

As 
in
Sec.~\ref{sec:TypicalPureStates:Derivation},
we define an ensemble of normalized 
random vectors $|\psi\rangle\in\hr$ via
\begin{eqnarray}
|\psi\rangle=
{\cal{N}}\, Q \sum_{n=1}^N z_n \, |\chi_n\rangle
\ ,
\label{a1app}
\end{eqnarray}
where the $z_n$ are complex numbers,
whose real and imaginary parts are 
given by independent, Gaussian distributed 
random variables of zero mean and 
unit variance.
Moreover, $Q$ is a linear operator on $\hr$,
which for the moment may still be 
(practically) arbitrary
(in particular, $Q$ need not be Hermitian);
for the ensemble under study in the main text,
e.g.,
we have
\begin{equation}
\label{a14}
	Q = P + y P A P \,,
\end{equation}
cf.\ Eq.~\eqref{eq:Q}.
Finally, ${\cal{N}}$ is a normalization 
constant so that 
$\langle\psi|\psi\rangle =1$.

A key property of the random vector ensemble 
in (\ref{a1app}) is its invariance under arbitrary 
unitary transformations of the basis 
$|\chi_n\rangle$ of $\hr$
(all statistical properties remain unchanged).
In other words, the basis can be chosen arbitrarily.
This is of particular interest when numerically
sampling random vectors according to (\ref{a1app}),
since any single-particle product basis 
will do the job.
In our numerical explorations, we always employed
such a ``computational basis'' $|\chi_n\rangle$.

Given some (non-zero) $Q$, one readily 
verifies that
\begin{eqnarray}
\rho:=Q\,Q^\dagger/\tr\{Q\,Q^\dagger\}
\ .
\label{a2app}
\end{eqnarray}
is Hermitian, positive semidefinite, and of unit 
trace, i.e., a well-defined density operator.
While $Q$ was until now (practically) 
arbitrary, we henceforth restrict 
ourselves to $Q$'s so
that $\rho$ in (\ref{a2app}) amounts to a mixed 
state of small purity, i.e., we require that
\begin{eqnarray}
\pu \ll 1
\ .
\label{a3}
\end{eqnarray}
(See also
Sec.~\ref{sec:TypicalPureStates:Derivation}, where 
Eq.~\eqref{a3} was established for the ensemble~\eqref{a2app} 
with $Q$ from~\eqref{a14}.)

Incidentally, a convenient way to numerically check
(\ref{a3}) is as follows:
Similarly as before, we sample random vectors $|\psi\rangle$
according to (\ref{a1app}), however we do not normalize them 
but rather set ${\cal{N}}=1$ in (\ref{a1app}).
Hence, the quantity $x:=\langle \psi|\psi\rangle$ 
is now a random variable. 
Likewise, when independently sampling
two random vectors, say $|\phi\rangle$ and $|\psi\rangle$,
according to (\ref{a1app}) with ${\cal{N}}=1$,
the quantity  $y:=|\langle \phi | \psi\rangle|^2$ 
will be another random variable.
Denoting by ${\mathbb E}[x]$ and ${\mathbb E}[y]$ the 
expectation values 
of $x$ and $y$,
one readily finds by means of the general framework 
from Ref. \cite{rei20} 
that $\pu={\mathbb E}[y]/({\mathbb E}[x])^2$.
Under the assumption that the probability distributions
of both random variables $x$ and $y$ are reasonably
well-behaved, a decent order-of-magnitude estimate
of their expectation values 
and of $\pu$ can thus be obtained by means of
only a quite small number of random vectors.

Next we consider, as in the main text, 
the pure states  $|\psi\rangle$
from (\ref{a1app}) as initial conditions
at time $t=0$,
which then evolve according to the 
Schr\"odinger equation 
$i\,\mbox{d} |\psi(t)\rangle/ \mbox{d} t=H|\psi(t)\rangle$, 
implying $|\psi(t)\rangle=U_t |\psi\rangle$
with $U_t:=e^{-iHt}$,
and resulting in expectation values 
$\langle O\rangle_{\! \psi(t)}:=\langle \psi (t)|O| \psi (t)\rangle$
for any given observable (Hermitian operator) $O$.
Likewise, the initial state $\rho$
in (\ref{a2}) is governed by the von
Neumann equation, implying
$\rho(t):=U_t\rho U_t^\dagger$.
Assuming that $\rho$ furthermore
obeys (\ref{a3}), 
the main finding (for our purposes) from
Ref.~\cite{rei20} then consist in the prediction that
the approximation
\begin{eqnarray}
\langle O\rangle_{\! \psi(t)}
\simeq
\tr\{\rho(t) O\}
\label{a4}
\end{eqnarray}
will be fulfilled with very high accuracy
for most normalized random vectors 
$|\psi \rangle$ in (\ref{a1app}). 
A quantitative version of this prediction in terms 
of the small parameter $\epsilon:=(\pu)^{1/3}$ 
can be obtained along similar lines as in 
Sec.~III.C of Ref.~\cite{rei18}, resulting in
\begin{eqnarray}
\prob\! \left(\, 
|\langle O\rangle_{\! \psi(t)} - \tr\{\rho(t) O\}|\geq \epsilon \DO
\,\right)
\leq 2 \epsilon
\ ,
\label{a5}
\end{eqnarray}
where $\prob(S)$ is the probability that the statement 
$S$ is true when randomly sampling an initial state
$|\psi\rangle$ according to (\ref{a1app}), and where
$\DO$ is the measurement range
(largest minus smallest eigenvalue) of the 
observable $O$.
In the setting from the main text,
the purity $\pu$ 
is usually exponentially small in the system size $L$ (see Sec.~\ref{sec:TypicalPureStates:Derivation}), 
hence also $\epsilon:=(\pu)^{1/3}$ in (\ref{a5})
is exponentially small in $L$.

Note that for any given $t$ there may still be
a small probability to sample an ``untypical'' 
initial state $|\psi \rangle$, for which  
(\ref{a4}) is a bad approximation.
Moreover, the set of all those untypical states 
may in general be different for different time points $t$.
Likewise, for any given observable $O$, the
set of untypical states $|\psi \rangle$ 
may in general be different.
Finally, also the following generalized
statement for any given observable $O$
can be readily shown along similar lines as  
in Sec.~17.4 of Ref.~\cite{bal19}:
Apart from a set of untypical initial states $|\psi\rangle$, 
whose probability is exponentially small in $L$,
the deviations in (\ref{a4}) are exponentially 
small in $L$ not only for an arbitrary but fixed
time point $t$ (as predicted by (\ref{a5})),
but even simultaneously for the vast majority of all time 
points within any preset time interval $[t_1,t_2]$,
where the relative measure of exceptional
time points is again exponentially small in $L$.

Apart from all those exponentially unlikely
exceptions, the main implication of
(\ref{a4}) is that initial states 
$|\psi \rangle$, randomly sampled according 
to (\ref{a1app}), are very likely to exhibit very 
similar expectation values 
at the initial time $t=0$ and also
at any later time $t>0$, a property which 
was originally discovered and 
named {\em dynamical typicality}
in Ref. \cite{bar09}.

\subsection{Equilibration and thermalization}
\label{app112}

Yet another relevant result from
Ref. \cite{rei20} is as follows:
Similarly as in the main text, 
we say that the system exhibits 
equilibration if the expectation value
$\langle O\rangle_{\! \psi(t)}$
remains very close to some constant 
``reference value'' 
for the vast majority of all sufficiently 
large times $t$, i.e., 
after initial transients 
(relaxation processes) 
have died out.
(We recall that a small fraction of exceptional,
arbitrarily large times $t$ is unavoidable due to 
quantum revival effects.)
On the other hand, the question of thermalization, 
i.e., whether or not this constant reference value 
is (nearly) equal to the pertinent thermal 
expectation value, 
is disregarded for the time being.
According to (\ref{a4}), most initial
states $|\psi\rangle$ will thus exhibit 
equilibration if the mixed state
$\rho(t)$ exhibits equilibration.
The latter has been demonstrated 
in Refs. \cite{equil}
under quite weak conditions on the 
eigenvalues $E_n$
of $H$ and on the initial state 
$\rho(0)=\rho$:
Essentially, it is sufficient that
the energy differences $E_m-E_n$
do not coincide for too many
index pairs with $m\not= n$
[which is the case for any generic 
Hamiltonian $H$ \cite{sre99,tas98,equil}],
and that -- in the absence of degeneracies --
all eigenstates $|n\rangle$  
of $H$ are weakly populated,
i.e., $\langle n|\rho|n\rangle\ll 1$.
As detailed in Ref.~\cite{rei20},
the latter requirement is equivalent to 
$\sum_{n=1}^N(\langle n|\rho|n\rangle)^2 \ll 1$,
which in turn is once again guaranteed
under the very same precondition as is 
(\ref{a3}). 
In case of degeneracies, a generalization along the
lines of \cite{equil} is straightforward, but its detailed
elaboration goes beyond our present scope.

Returning to the question of thermalization,
and taking  for granted the above mentioned 
conditions for equilibration, it is thus sufficient to
show that the long-time average of the expectation
value on the right hand side of (\ref{a4}) is well 
approximated by the corresponding thermal 
expectation value.
In a first step, we therefore introduce 
$U_t:=e^{-iHt}$ and $\rho(t):=U_t\rho U_t^\dagger$
(see above (\ref{a4})) on the right
hand side of (\ref{a4}), and evaluate
the trace in terms of the eigenvalues $E_n$ and
eigenvectors $|n\rangle$ of $H$,
yielding
\begin{eqnarray}
\tr\{\rho(t) O\}
=
\sum_{m,n=1}^N \rho_{mn}\, O_{nm}\, e^{i(E_n-E_m)t}
\label{a6}
\end{eqnarray}
where $\rho_{mn}:=\langle m| \rho|n\rangle$ and 
$O_{nm}:=\langle n| O |m\rangle$.
In case of degeneracies, we can and will choose 
the eigenvectors $|n\rangle$ so that $O$ is 
diagonal in the corresponding eigenspaces 
of $H$.
Indicating the long-time average by an overline,
we thus can conclude that
\begin{eqnarray}
\overline{\tr\{\rho(t) O\}}
=
\sum_{n=1}^N \rho_{nn}\, O_{nn}
=\tr\{\overline{\rho}\,O\}
\ ,
\label{a7}
\end{eqnarray}
where $\overline{\rho}$ is the so-called diagonal ensemble 
or long-time average of $\rho(t)$,
\begin{eqnarray}
\overline{\rho} := \sum_{N=1}^N\rho_{nn}\, |n\rangle\langle n|
\ .
\label{a8}
\end{eqnarray}

The remaining task is to show that the long-time 
average in (\ref{a7}) is (approximately) equal to the
corresponding thermal equilibrium expectation 
value,
which was accomplished in Sec.~\ref{sec:TypicalPureStates:Derivation}.

Quantitatively, our conclusions are as follows:
Given that the purity $\tr\{\rho^2\}$ is exponentially 
small in the system size (see Sec.~\ref{sec:TypicalPureStates:Derivation}), 
the typical differences
$\langle O\rangle_{\! \psi(t)} - \tr\{\rho(t) O\}$
were found, as detailed below (\ref{a5}), 
to be exponentially small in the system size $L$
for the vast majority of all (sufficiently late) times $t$,
the exceptional $t$'s and $|\psi\rangle$'s being 
exponentially rare in $L$.
Analogous conclusions can be shown \cite{equil}
to apply to the differences
$\tr\{\rho(t) O\}-\tr\{\overline{\rho}\,O\}$.
Finally, the difference
$\overline{\Delta O} := \tr\{\bar\rho O\} - \Omc$
was found to obey 
$\overline{\Delta O}^2 \leq c \Delta_O^2$
(see below Eq. (\mref{zz5})),
where $c$ is a constant and  $\Delta_O^{\,2}$ 
the wETH characteristic from (\mref{zz3}).
Quantitatively, pertinent previous investigations in 
Refs.~\cite{bir10,alb15,mor16,iyo17,mor18,yos18,kuw20,kuw20a}
suggest that $\Delta_O^{\,2}$  
decreases (approximately) like $1/L$ 
with $L$.
Altogether, $\langle O\rangle_{\!\psi(t)}-\Omc$ 
is thus predicted to decrease 
as $L^{-1/2}$ for the vast majority of all (sufficiently late) 
times $t$ and initial states $|\psi\rangle$,
the exceptional $t$'s and $|\psi\rangle$'s 
being exponentially rare in $L$.

\section{Numerical implementation}
\label{app:B}

Overall, (\ref{a1app}) together with~(\ref{a14}) amount to
an explicit procedure of how to generate
initial states $|\psi\rangle$ 
with a 
non-equilibrium
expectation value of some given (local) observable $A$.
The vast majority of those expectation values will
be very close to the value on the right
hand side of (\mref{a22x}), which in turn 
can be adjusted by properly choosing the 
parameter $y$.
A numerical implementation of this
procedure is straightforward in principle
(see also the discussion below (\ref{a1app})).
In practice, the required projector $P$ 
onto the energy shell $\hat \hr$ cannot be determined 
without diagonalizing the Hamiltonian $H$,
thus limiting the numerics to relatively 
small systems.
The main objective of this appendix is to circumvent
such a diagonalization.

\subsection{Setup}

A well-established way to numerically overcome 
this problem \cite{filter}
is to approximate the projector $P$
onto the energy shell $\hat \hr$ by an 
``energy-filter'' of the general form
\begin{eqnarray}
P_\nu:=e^{-\gamma(H-\lambda)^{2\nu}}
\label{a25}
\end{eqnarray}
with $\nu\in\NN$
and suitably chosen parameters
$\gamma$ and $\lambda$.
Indeed, for sufficiently large $\nu$ and appropriate
values of $\gamma$ and $\lambda$, the effect 
of $P_{\nu}$ approximates that of $P$ arbitrarily
well for any eigenvector $|n\rangle$ of $H$
and thus for any $|\phi\rangle\in\hr$.
Similarly, one may expect that $\nu=1$
still amounts to an acceptable compromise 
between a reasonable approximation of $P$
and numerical feasibility \cite{filter}.
From now on, we thus restrict ourselves to 
such ``Gaussian energy filters'' (hence the subscript ``$g$'')
of the form
\begin{eqnarray}
\Pg :=e^{-\gamma (H-\lambda)^{2}}
\ .
\label{a26}
\end{eqnarray}

As said in the main text, our actual numerical
implementation of how the operator $\Pg$ acts
on any given state $|\phi\rangle\in\hr$ is based
on imaginary time evolution methods in combination 
with Suzuki-Trotter product expansion techniques \cite{rae04}.

Since the fundamental property $P^2=P$ of a genuine 
projector is no longer rigorously satisfied by $\Pg$
from (\ref{a26}),
the appropriate way of how to replace $P$ by $\Pg$ 
in (\ref{a14}) is not immediately obvious.
For instance, (\ref{a14}) is equivalent to
$Q=P(1+yA)P$, but when replacing $P$ by $\Pg$,
two different descendants of the two originally 
identical $Q$'s are obtained.
To us, the most natural modification of (\ref{a14})
seems to be 
\begin{eqnarray}
Q=\Pg (1+yA)\Pg 
\ .
\label{a27}
\end{eqnarray}

It should be emphasized that in the end it will turn 
out not to be very important how closely (\ref{a26})
and (\ref{a27}) approximate their original
counterparts
(and likewise for similar further approximations later on).
The reason is that we finally will obtain a new
(numerical) way in its own right of how to generate
initial states $|\psi\rangle$ which exhibit
non-equilibrium
expectation values of  $A$
and thermalization in the long run.

Given (\ref{a27}), the sampling of random states
$|\psi\rangle$ and the density operator $\rho$
are again determined by (\ref{a1app}) and (\ref{a2app}),
the condition (\ref{a3}) can again be numerically
examined as described below (\ref{a3}),
and also the subsequent steps (\ref{a4})-(\ref{a8})
remain valid. 

Practically, results for many different parameter 
values $y$ in (\ref{a27}) can be conveniently 
produced in one numerical run as follows.
The basic idea is to divide (\ref{a1app}) into
several substeps by numerically 
generating the (not normalized) random 
vectors (see also discussion below (\ref{a1app}))
\begin{eqnarray}
|\phi\rangle & := & \sum_{n=1}^N z_n \, |\chi_n\rangle
\ ,
\label{y2}
\\
|\psi_1\rangle & := & P_g^2|\phi\rangle
\ ,
\label{y3}
\\
|\psi_2\rangle & := & P_gA\Pg |\phi\rangle
\ .
\label{y4}
\end{eqnarray}
The normalized vector $|\psi\rangle$ in (\ref{a1app}) with $Q$ from
(\ref{a27}) is thus recovered via
\begin{eqnarray}
|\psi\rangle & := & {\cal{N}} (|\psi_1\rangle + y |\psi_2\rangle)
\ .
\label{y5}
\end{eqnarray}
However, instead of time evolving $|\psi\rangle$ 
and then evaluating the expectation values of some
observables of interest, one can also time evolve
$|\psi_1\rangle$ and $|\psi_2\rangle$ separately,
and only then evaluate the expectation values,
and analogously for the estimate of $\pu$
described below (\ref{a3}).
For instance, to determine the normalization 
constant ${\cal N}$ for arbitrary $y$ values, 
one only needs the three ($y$-independent)
quantities $\langle\psi_1|\psi_1\rangle$,
$\langle\psi_1|\psi_2\rangle$, and
$\langle\psi_2|\psi_2\rangle$,
and analogously for any expectation value.
In particular, one can thus numerically 
adapt $y$ so that
the initial expectation value of $A$ assumes
some preset value (provided it can be realized 
at all by some value of $y$),
and one can evaluate the behavior of 
several different such initial values 
in one numerical run.

The next problem is how to appropriately 
define the thermal equilibrium expectation 
value of $O$, with which the long-time average
in (\ref{a7}) must be compared in order to 
decide whether or not thermalization 
takes place.
The usual and natural solution is to choose once again
some suitable energy window 
$\Imc:=[E-\Delta,E]$ 
so that the corresponding microcanonical
ensemble $\rhomic$ (see also main text) 
imitates the energy distribution of the actual 
$\rho$ in (\ref{a2app}) as closely as possible. 
More precisely, $E$ and $\Delta$
must be chosen so that the mean and 
variance of the energy distributions agree,
i.e., $\tr\{\rhomic H\}=\tr\{\rho H\}$ and 
$\tr\{\rhomic H^2\}=\tr\{\rho H^2\}$
(note that $\tr\{\rho H^k\}=\tr\{\rho(t) H^k\}
=\tr\{\overline{\rho} H^k\}$
for any $k\in\NN$ according to 
(\ref{a6}) and (\ref{a7})).
However, to determine such an ensemble
$\rhomic$ is again numerically inconvenient
for the same reasons as those mentioned 
at the beginning of this section.

The solution of this problem is based on the 
fact that a so-called equivalence-of-ensembles
property has been derived in Refs. \cite{kuw20a,tas18,bra15}
basically under the same preconditions as those
which are required in the proofs of the wETH,
and which we
thus tacitly take for granted anyway
(see also main text).
Essentially, this means that the above specified
microcanonical ensemble $\rhomic$ can be
replaced by other ensembles of the general ``diagonal form''
(\ref{a8}), provided that the level populations $\rho_{nn}$
depend sufficiently smoothly on the energies $E_n$,
and, as before, the mean and variance of the energy
distribution are close to those of the actual 
ensemble from (\ref{a2app}).

Obviously, a natural candidate of this kind is obtained
by setting $y=0$ in (\ref{a27}) and then 
evaluating (\ref{a2app}).
Indeed, similarly as
observed around Eq.~(\mref{a22x}) in the main text,
it turns out
that only local observables $A$ are of actual interest,
for which the mean and variance of the 
energy distribution hardly change with 
$y$ in (\ref{a27}).
Altogether, it is thus justified to approximate the thermal 
equilibrium expectation $\Omc:=\tr\{\rhomic O\}$ by
\begin{eqnarray}
\Omc= \tr\{\rhoth O\}
\ ,
\label{a28}
\end{eqnarray}
where the  ``thermal reference ensemble'' (index ``th'') 
is defined as
\begin{eqnarray}
\rhoth:=(\Pg )^4/\tr\{(\Pg )^4\}
\ .
\label{a29}
\end{eqnarray}
see also Appendix~\ref{app:C} below.
Note that the above mentioned equivalence
of ensembles and the concomitant approximation 
(\ref{a28}) are predicted to become exact for 
asymptotically large systems, 
while for finite systems it is reasonable to expect 
that the remaining difference between long-time 
average and thermal value will usually be 
even smaller when employing the auxiliary thermal 
ensemble $\rhoth$ from (\ref{a29}) instead of its 
microcanonical counterpart $\rhomic$.

Numerically, the expectation value on the right
hand side of (\ref{a28}) can once again be readily approximated
by choosing $y=0$ in the procedure described below (\ref{y5}).

Note that all traces of an energy interval (or energy shell)
have now disappeared. Hence the basic requirement that
all considered states $|\psi\rangle$ must exhibit a narrow
energy distribution (see main text) has now to 
be (numerically) verified 
by choosing $y=0$ and evaluating the variance
$\langle H^2\rangle_{\psi} - \langle H\rangle_{\psi}^2$
along the lines described below (\ref{y5}).

In this context it is also noteworthy 
that the mean and variance of the energy
distribution for the ensemble in (\ref{a29})
are only well approximated by $\lambda$ and 
$1/8\gamma$ according to (\ref{a26}) under the 
condition that the level density of $H$ can
be considered as approximately constant within
a neighborhood of $\lambda$ on the order of 
$\gamma^{-1/2}$. 
Otherwise, the relation of the
parameters $\lambda$ and $\gamma$ to
the mean and variance of the energy 
distribution is non-trivial and must be
determined numerically.

\subsection{Dynamical typicality}

Following Appendix~\ref{app:A},
dynamical typicality for the present $\rho$ from~\eqref{a2app} with $Q$ from \eqref{a27}
(and $y \neq 0$)
is established if we can show $\tr\{\rho^2\}$ to be small,
cf.\ Eq.~\eqref{a3}.
This can be achieved by a tedious, but straightforward calculation,
whose steps parallel those described in the main text,
see Sec.~\ref{sec:TypicalPureStates:Derivation}.
It results in the bound
\begin{eqnarray}
\tr\{\rho^2\} \leq \pmax\, \frac{1+6y^2+4m_3y^3+m_4y^4}{(1+\tilde m_2 y^2)^2} \,,
\label{a43}
\end{eqnarray}
where $m_k := \tr\{ \rhoth A^k_r \}$ are the equivalent of~(\mref{a1}) with $\rhoth$ from~\eqref{a29} in lieu of $\rhomic$
and $A_r$ is the rescaled variant of $A$ such that $m_1 = 0$ and $m_2 = 1$.
Furthermore, $\tilde m_2 := \tr\{ (\rhoth^{1/2} A_r)^2 \}$ (implying $0 \leq \tilde m_2 \leq 1$) and
\begin{equation}
\label{eq:pmax}
	p_{\max} := \max_n \rho_{nn} \,.
\end{equation}
For any reasonable energy window, $\gamma$ in~\eqref{a26} will be much smaller 
than the squared energy level density around the target energy $\lambda$.
In particular, assuming a natural scaling of the energy window's width with $L^{1/2}$, 
$\gamma$ will decrease as $L^{-1}$,
hence $p_{\max}$ from~\eqref{eq:pmax} will be exponentially small in $L$.
Moreover, $m_3$ and $m_4$ will usually be (at most) on the order of unity 
(see also the discussion below~(\mref{a4}) in the main text),
such that
$\tr\{\rho^2\}$ will be exponentially small in $L$ according to~\eqref{a43}.

\subsection{Thermalization}
\label{app:B:Thermalization}

Finally,
to demonstrate thermalization,
we need to show again that
\begin{equation}
\overline{\Delta O} := \tr\{\bar\rho O\} - \Omc
= \sum_{n=1}^N \rho_{nn} (O_{nn}-\Omc)
\label{a19}
\end{equation}
is negligibly small,
cf.\ above Eq.~(\mref{zz4}).
In the definition
$\Omc:=\tr\{\rhomic O\}$ there, the microcanonical
ensemble $\rhomic$ has now to be chosen, as usual,
so that the ``true'' system energy $E:=\tr\{\rho H\}$
is correctly reproduced by $\tr\{\rhomic H\}$.
Furthermore,
another auxiliary microcanonical ensemble
$\rhomic'$ is
introduced
so that $\tr\{\rhomic' H\}$ agrees
with the energy $E':=\tr\{\Pc H\}$ of the ensemble $\Pc$ 
from 
(\ref{a29}),
while the concomitant expectation values 
are denoted as $\Omc':=\tr\{\rhomic' O\}$.
This allows us to rewrite $\overline{\Delta O}$ from~\eqref{a19} as
\begin{eqnarray}
\overline{\Delta O} & = & X_1+X_2
\ ,
\label{a49a}
\\
X_1 & := &  \sum_{n=1}^N \rho_{nn} (O_{nn}-\Omc')
\ ,
\label{a49b}
\\
X_2 & := & \Omc'-\Omc \,.
\label{a49c}
\end{eqnarray}
A similar decomposition for the canonical ensemble was performed 
in Eqs.~(\mref{zz8})--(\mref{zz10}) in the main text.
In the following, we will argue that $X_1$ is negligibly small for 
the Gaussian energy-filter approximation of the microcanonical ensemble, 
Eqs.~\eqref{a2app} and~\eqref{a27}.
The smallness of $X_2$ from~\eqref{a49c} will be established in 
Appendix~\ref{app:C} as this quantity is essentially equivalent to~(\mref{zz10}).

Turning to $X_1$ from~\eqref{a49b},
the crucial
observation is 
that $\rho_{nn}:=\langle n| \rho|n\rangle$
(see below (\ref{a6})) can be rewritten
by means of
(\ref{a2app}), (\ref{a27}), and (\ref{a29})
(see also around (\ref{a43}))
in the form
\begin{eqnarray}
\rho_{nn} & = & p_n^{1/2} r_n
\ ,
\label{a50}
\\
r_n & := & \frac{\langle n| D |n\rangle}{1+\tilde m_2 y^2}
\ ,
\label{a50a}
\\
p_n & := & \bra{n} \rhoth \ket{n}
\ ,
\\
D & := & (1 + y A_r) \rhoth^{1/2} (1 + y A_r)
\ .
\end{eqnarray}
In fact,
the following considerations will apply to rather general thermal reference ensembles of the form
\begin{eqnarray}
\Pc :=\sum_{n=1}^N p_n \, |n\rangle\langle n|
\ ,
\label{a30}
\end{eqnarray}
where the $p_n$ must satisfy the properties
\begin{eqnarray}
p_n & \geq & 0 \ ,
\label{a31}
\\
\sum_{n=1}^N p_n & = & 1
\ ,
\label{a32}
\end{eqnarray}
which obviously includes the Gaussian energy filters~\eqref{a29}
as well as the microcanonical ensemble $\rhomic$ from the main 
text (see beginning of
Sec.~\ref{sec:Preliminaries:TypClustWETH})
and the canonical (Gibbs) ensemble from~(\mref{zz11})--(\mref{zz13}).

Utilizing~\eqref{a50},
we can rewrite (\ref{a49b}) as
\begin{eqnarray}
X_1 = \sum_{n=1}^N r_n \left[p_n^{1/2}(O_{nn}-\Omc')\right]
\ .
\label{a51}
\end{eqnarray}
Along the same lines as 
around~(\mref{zz14})--(\mref{zz16}),
one can infer from (\ref{a50a}) and (\ref{a51}) that
\begin{eqnarray}
X_1^2 & \leq & c'' \, \Delta_O^{\prime \, 2}
\ ,
\label{a52}
\\
c'' & := & 
\frac{\sum_{n=1}^N \langle n| D |n\rangle^2}{(1+\tilde m_2 y^2)^2}
\leq \frac{\tr\{D^2\}}{(1+\tilde m_2 y^2)^2}
\ ,
\label{a53}
\\
\Delta_O^{\prime \, 2} & := & \sum_{n=1}^N p_n (O_{nn}-\Omc')^2
\ .
\label{a54}
\end{eqnarray}
Note that $\Delta_O^{\prime \,2}$ in~\eqref{a54} is the same quantity as in~(\mref{zz16})
if we assume the general form~\eqref{a30} for the thermal ensemble $\rhoth$.
Similarly as around
(\ref{a43}), it follows that
\begin{eqnarray}
c'' \leq 
\frac{ 1+6y^2+4m_3y^3+m_4y^4 }{ (1+\tilde m_2 y^2)^2 }
\leq  \ord (1)
\ .
\label{a55}
\end{eqnarray}

Finally, we provide more detailed arguments for the smallness 
of $\Delta_O^{\prime \, 2}$ from~\eqref{a54}, see also the discussion below~(\mref{zz16}).
Recalling that we are dealing with Hamiltonians $H$ and observables $O$ satisfying the wETH,
the matrix elements $O_{nn}$ can be 
essentially considered as pseudo-random numbers, 
which exhibit, as a function of the corresponding energies
$E_n$, some reasonably well-defined running (or local)
average so that the concomitant (local) variance
becomes negligibly small for large systems sizes $L$ 
[cf.\ Eq.~(\mref{zz3})].
Accordingly, also $\Delta_O^{\prime \, 2}$ in (\ref{a54}) becomes
negligibly small provided the weights $p_n$ are
mainly concentrated within some sufficiently small 
energy interval, within which the running average
of the $O_{nn}$ hardly changes
(and thus is necessarily always close to $\Omc'$).
Moreover, the $p_n$'s must not be correlated
with the pseudo-random fluctuations of the 
$O_{nn}$'s about their running average. 
The latter requirement will, in particular, be
fulfilled if the weights $p_n$ change sufficiently
slowly as a function of the energies $E_n$ 
(as is the case for the Gaussian filters
from  (\ref{a26}), (\ref{a29}) and for the
canonical ensembles from (\mref{zz11})--(\mref{zz13})),
with the possible exception of discontinuous
jumps, which are sufficiently far apart
from each other (as is the case for the
microcanonical ensembles from 
the main text).
Taking the absence of such correlations for granted,
and denoting the mean energy, as above (\ref{a49a}), 
by $E':=\tr\{\Pc H\}$,
our main additional requirement 
on $\Pc$ is thus that the corresponding energy spread
(standard deviation)
\begin{eqnarray}
\sigma':=\sqrt{\tr\{\Pc(H-E')^2\}}
\label{a57}
\end{eqnarray}
must be very small compared to the typical system 
energies $E'$ themselves.
Since those energies generically exhibit a linear dependence
on the system size $L$, we thus require that $\sigma'$ 
grows at most sublinearly with $L$.
Again, the latter condition is generically fulfilled, in particular,
for our usual examples (microcanonical ensemble, 
Gaussian energy filters, canonical ensembles).
Altogether, the task to identify and verify conditions
under which $\Delta_O^{\prime \, 2}$ from (\ref{a54}) is small
can thus be considered as settled,
which in turn implies that $X_1$ in~\eqref{a49b} 
and likewise in~(\mref{zz9}) are
small.

\section{Bound on differences of thermal expectation values}
\label{app:C}

In this appendix,
we argue that $X_2$ from~(\mref{zz10}) and~\eqref{a49c} are small quantities,
assuming the general form~\eqref{a30} for the thermal reference ensemble $\rhoth$
(e.g., the canonical ensemble~(\mref{zz11})--(\mref{zz13}) or the Gaussian filters (\ref{a26}), (\ref{a29})).

Besides the subextensive scaling of $\sigma'$ from~\eqref{a57},
we also require
that the so-called fourth 
central moment $ \tr\{\Pc\, (H-E')^4\}$ is still roughly 
comparable in order of magnitude to the square of the
variance (second central moment), i.e.
\begin{eqnarray}
\tr\{\Pc (H-E')^4\} = \ord[ (\sigma')^4 ]
\ .
\label{a59}
\end{eqnarray}
This requirement is very weak since it only excludes
energy distributions with very slowly decaying tails.
For instance, for the canonical ensemble (\mref{zz11})--(\mref{zz13})
the energy distribution is well-known to be 
approximately Gaussian, 
i.e., $ \tr\{\Pc(H-E')^4\}$
is close to $3(\sigma')^4$, provided the
temperature $\beta^{-1}$ is not too close to zero.
Analogous considerations generically
also apply to the energy filters from (\ref{a26}) 
and to the microcanonical setup from 
the first part of the main text, provided the respective model 
parameters are chosen within reasonable limits.
In other words, the requirement (\ref{a59})
is usually fulfilled automatically.

Denoting, similarly as above (\ref{a49a}) and 
around (\ref{a57}), 
the mean and variance of the energy distribution 
of $\rho$,
which may either be of the dynamical-typicality form~\eqref{a2app} 
with $Q$ from~\eqref{a27} or of the pre-quench Gibbs form~(\mref{zz7}),
by
\begin{eqnarray}
E & := & \tr\{\rho H\}
\ ,
\label{a60}
\\
\sigma^2 & := & \tr\{\rho (H-E)^2\}
\ ,
\label{a61}
\end{eqnarray} 
and defining $\Delta:=E-E'$,
one readily confirms that
\begin{eqnarray}
W  & := & \tr\{\rho (H-E')^2\}
=
\tr\{\rho [(H-E)+\Delta]^2\}
\nonumber
\\
& =  &  
\tr\{\rho (H-E)^2\} +
2\tr\{\rho (H-E)\}\Delta+\Delta^2
\nonumber
\\
& = &
\sigma^2+(E-E')^2
\ .
\label{a62}
\end{eqnarray}
Recalling that $|n\rangle$ and $E_n$ are the eigenvectors and
eigenvalues of $H$ and that $\rho_{nn}:=\langle n|\rho| n \rangle$
it follows that
\begin{eqnarray}
W  = \sum_{n=1}^N \rho_{nn}\, (E_n-E')^2
\ .
\label{a63}
\end{eqnarray}
Similarly as in~\eqref{a52} (Gaussian filter) or~(\mref{zz14}) (Gibbs ensemble),
this implies
\begin{eqnarray}
W^2 & \leq & \tilde c \, W'
\ ,
\label{a63a}
\end{eqnarray}
where $\tilde c$ is either $c''$ from~(\ref{a53}) or 
$c'$ from~(\mref{zz15}) and is thus again upper-bounded by an $L$-independent constant,
while $W' $ is defined as
\begin{eqnarray}
W'  := \sum_{n=1}^N p_n (E_n-E')^4
=  \tr\{\Pc (H-E')^4\}
\ ,
\label{a64a}
\end{eqnarray}
where the last step is based on similar 
arguments as below (\ref{a62}).
Observing $\tilde c \leq \mathcal{O}(1)$
and (\ref{a59}), we thus can conclude from
(\ref{a63}) that
\begin{eqnarray}
W  \leq \ord [(\sigma')^2]
\ ,
\label{a65a}
\end{eqnarray}
and with (\ref{a62}) that
\begin{eqnarray}
\sigma & \leq & \ord(\sigma')
\ ,
\label{a64}
\\
|E'-E| & \leq & \ord(\sigma')
\ .
\label{a65}
\end{eqnarray}

Since $\Pc$ exhibits a narrow energy distribution,
as detailed below (\ref{a57}), the same follows
for the energy distribution of $\rho$ according to
(\ref{a61}) and (\ref{a64}).
Given the precondition (\ref{a3}) for dynamical typicality 
is fulfilled (see main text and Appendix~\ref{app:B}),
this finally guarantees 
that also the vast majority of the random 
states $|\psi\rangle$ in (\ref{a1app}) exhibit a narrow 
energy distribution.
In other words, the basic requirement throughout 
the main text (see also below (\ref{a29}))
can thus be taken for granted.
All this is obviously also consistent with the common
wisdom that thermalization is in general not to be expected
in the absence of a narrow energy distribution.

Similarly to the discussion around (\ref{a57}),
the microcanonical expectation values of relevant 
observables are expected to only exhibit negligible
variations as long as the energy changes of 
the microcanonical ensemble are sublinear 
in the system size $L$.
We thus can infer from (\ref{a65}) that the difference $X_2$
of the microcanonical expectation values
in (\ref{a49c}) and~(\mref{zz10}) is negligible, i.e., our
demonstration of thermalization is complete.

Put differently, the approximation $\Omc=\Omc'$ 
can be considered as very well fulfilled.
Finally, and similarly as above (\ref{a28}), 
the approximation $\Omc'=\tr\{\Pc O\}$ can 
be taken for granted by the quite general equivalence 
of ensembles as detailed in Refs.
\cite{kuw20a,tas18,bra15}.
Altogether, we thus recover the approximation
\begin{eqnarray}
\Omc= \tr\{\Pc O\}
\ ,
\label{a44}
\end{eqnarray}
amounting to a more detailed justification 
and generalization of (\ref{a28}).



\begin{thebibliography}{70}

\bibitem{neu29}
J. von Neumann,
Beweis des Ergodensatzes und des H-Theorems in der neuen Mechanik,
Z. Phys. {\bf 57}, 30-70 (1929)
[English translation by R. Tumulka,
Proof of the ergodic theorem and the H-theorem in quantum mechanics,
Eur. Phys. J. H {\bf 35}, 201-237 (2010)].

\bibitem{deu91}
J. M. Deutsch, 
Quantum statistical mechanics in a closed system,
Phys. Rev. A {\bf 43}, 2046-2049 (1991).

\bibitem{sre96}
M. Srednicki, 
Thermal fluctuations in quantized chaotic systems, 
J. Phys. A {\bf 29}, L75 (1996).

\bibitem{tas98}
H. Tasaki, 
From quantum dynamics to the canonical distribution:
general picture and rigorous example,
Phys. Rev. Lett. {\bf 80}, 1373-1376 (1998).

\bibitem{sre94}
M. Srednicki, 
Chaos and quantum thermalization,
Phys. Rev. E {\bf 50}, 888-901 (1994).

\bibitem{rig08}
M. Rigol, V. Dunjko, and M. Olshanii,
Thermalization and its mechanism for generic
isolated quantum systems,
Nature {\bf 452}, 854-858 (2008).

\bibitem{tas16}
H. Tasaki,
Typicality of Thermal Equilibrium and Thermalization 
in Isolated Macroscopic Quantum Systems,
J. Stat. Phys. {\bf 163}, 937 (2016).

\bibitem{ued20}
M. Ueda,
Quantum equilibration, thermalization and prethermalization in ultracold atoms, 
Nat. Rev. Phys. {\bf 2}, 669 (2020).

\bibitem{dal16}
L. D'Alessio, Y. Kafri, A. Polkovnikov, and M. Rigol,
From Quantum Chaos and Eigenstate Thermalization
to Statistical Mechanics and Thermodynamics,
Adv. Phys.  {\bf 65}, 239 (2016).

\bibitem{gog16}
C. Gogolin and J. Eisert,
Equilibration, thermalization, and the emergence
of statistical mechanics in closed quantum systems,
Rep. Prog. Phys. {\bf 79}, 056001 (2016).

\bibitem{mor18}
T. Mori, T. N. Ikeda, E. Kaminishi, and M. Ueda,
Thermalization and prethermalization 
in isolated quantum systems: a theoretical overview,
J. Phys. B  {\bf 51}, 112001 (2018).

\bibitem{bir10}
G. Biroli, C. Kollath, and A. L\"auchli,
Effect of rare fluctuations on the thermalization of isolated quantum systems.
Phys. Rev. Lett. {\bf 105}, 250401 (2010).

\bibitem{mor16}
T. Mori,
Weak eigenstate thermalization with large deviation bound,
arXiv:1609.09776 (2016).

\bibitem{iyo17}
E. Iyoda, K. Kaneko, and T. Sagawa,
Fluctuation Theorem for Many-Body Pure Quantum States,
Phys. Rev. Lett. {\bf 119}, 100601 (2017).

\bibitem{kuw20}
T. Kuwahara and K. Saito,
Eigenstate thermalization from clustering property of correlations,
Phys. Rev. Lett. {\bf 124}, 200604 (2020).

\bibitem{kuw20a}
T. Kuwahara and K. Saito,
Gaussian concentration bound and ensemble equivalence in generic quantum 
many-body systems including long-range interactions,
Ann. Phys. {\bf 421}, 168278 (2020).

\bibitem{llo88}
S. Lloyd, Ph.D. thesis, The Rockefeller University, 1988,
Chapter 3, arXiv:1307.0378.

\bibitem{gol06}
S. Goldstein, J. L. Lebowitz, R. Tumulka, and N. Zhang\`{\i},
Canonical typicality,
Phys. Rev. Lett. {\bf 96}, 050403 (2006).

\bibitem{pop06}
S. Popescu, A. J. Short, and A. Winter,
Entanglement and the foundations of statistical mechanics,
Nat. Phys. {\bf 2}, 754-758 (2006).

\bibitem{mul15}
M. P. M\"uller, E. Adlam, L. Masanes, and N. Wiebe,
Thermalization and canonical typicality in translation-invariant quantum lattice systems,
Commun. Math. Phys. {\bf 340}, 499 (2015).

\bibitem{ess16}
F. H. L. Essler and M. Fagotti,
Quench dynamics and relaxation in isolated integrable quantum spin chains,
J. Stat. Mech. 064002 (2016).

\bibitem{far17}
T. Farrelly, F. G. S. L.  Brand\~ao, and M. Cramer,
Thermalization and return to equilibrium on finite quantum lattice systems,
Phys. Rev. Lett. {\bf 118}, 140601 (2017).

\bibitem{doy17}
B. Doyon,
Thermalization and pseudolocality in extended quantum systems,
Commun. Math. Phys. {\bf 351}, 155 (2015).

\bibitem{sot14}
S. Sotiriadis and P. Calabrese,
Validity of the GGE for quantum quenches from interacting to noninteracting models,
J. Stat. Mech. P07024 (2014).

\bibitem{mur19}
C. Murthy and M. Srednicki,
Relaxation to Gaussian and generalized Gibbs states in systems of particles with quadratic Hamiltonians,
Phys. Rev. E {\bf 100}, 012146 (2019).

\bibitem{glu19}
M. Gluza, J. Eisert, and T. Farrelly,
Equilibration towards generalized Gibbs ensembles for non-interacting systems,
SciPost {\bf 7}, 038 (2019).

\bibitem{wei97}
S. Weinberg, 
{What is Quantum Field Theory, and What Did We Think It Is?},
arXiv:hep-th/9702027 (1997).

\bibitem{ara69}
H. Araki,
Gibbs states of a one dimensional quantum lattice,
Commun. Math. Phys. {\bf 14}, 120 (1969).

\bibitem{par82}
Y. M. Park,
The cluster expansion for classical and quantum lattice systems,
J. Stat. Phys. {\bf 27}, 553 (1982).

\bibitem{par95}
Y. M. Park and H. J. Yoo,
Uniqueness and clustering properties of Gibbs states 
for classical and quantum unbounded spin systems,
J. Stat. Phys. {\bf 80}, 223 (1995).

\bibitem{kli14}
M. Kliesch, C. Gogolin, M. J. Kastoryano, A. Riera, and J. Eisert,
Locality of temperature,
Phys. Rev. X {\bf 4}, 031019 (2014).

\bibitem{fro15}
J. Fr\"ohlich and D. Ueltschi,
Some properties of correlations of quantum lattice 
systems in thermal equilibrium,
J. Math. Phys. (N.Y.) {\bf 56}, 053302 (2015).

\bibitem{tas18}
H. Tasaki,
On the local equivalence between the canonical and the microcanonical 
ensembles for quantum spin systems,
J. Stat. Phys. {\bf 172}, 905 (2018).

\bibitem{cas11}
A. C. Cassidy, C. W. Clark, and M. Rigol,
Generalized thermalization in an integrable lattice system,
Phys. Rev. Lett. {\bf 106}, 140405 (2011).

\bibitem{bar09}
C. Bartsch and J. Gemmer, 
Dynamical typicality of quantum expectation values,
Phys. Rev. Lett. {\bf 102}, 110403 (2009).

\bibitem{rei20}
P. Reimann and J. Gemmer,
Why are macroscopic experiments reproducible? 
Imitating the behavior of an ensemble by single pure states,
Physica A {\bf 552}, 121840 (2020).

\bibitem{vid16}
L. Vidmar and M. Rigol,
Generalized Gibbs ensemble in integrable lattice models,
J. Stat. Mech. 064007 (2016).

\bibitem{filter}
R. Steinigeweg, A. Khodja, H. Niemeyer, C. Gogolin, and J. Gemmer,
Pushing the limits of the eigenstate thermalization hypothesis towards mesoscopic quantum systems,
Phys. Rev. Lett. {\bf 112}, 130403 (2014).

\bibitem{rae04}
H. De Raedt and K. Michielsen, Computational Methods for Simulating Quantum Computers, arXiv:quant-ph/0406210 (2004).

\bibitem{rei18a}
P. Reimann,
Dynamical typicality approach to eigenstate thermalization,
Phys. Rev. Lett. {\bf 120}, 230601 (2018).

\bibitem{equil}
P. Reimann, 
Foundation of statistical mechanics under 
experimentally realistic conditions,
Phys. Rev. Lett. {\bf 101}, 190403  (2008);
N. Linden, S. Popescu, A. J. Short, and A. Winter, 
Quantum mechanical evolution towards equilibrium,
Phys. Rev.  E {\bf 79}, 061103 (2009);
A. J. Short, 
Equilibration of quantum systems and subsystems,
New J. Phys. {\bf 13}, 053009  (2011);
P. Reimann and M. Kastner, 
Equilibration of macroscopic quantum systems,
New J. Phys. {\bf 14}, 043020 (2012);
A. J. Short and T. C. Farrelly,
Quantum equilibration in finite time,
New J. Phys. {\bf 14}, 013063 (2012);
B. N. Balz and P. Reimann, 
Equilibration of isolated many-body quantum systems with 
respect to general distinguishability measures, 
Phys. Rev. E {\bf 93}, 062107 (2016).

\bibitem{sre99}
M. Srednicki, 
The approach to thermal equilibrium in quantized chaotic systems,
J. Phys. A: Math. Gen {\bf 32}, 1163 (1999).

\bibitem{ric20}
J. Richter, A. Dymarsky, R. Steinigeweg, and J. Gemmer,
Eigenstate thermalization hypothesis beyond standard indicators: 
Emergence of random-matrix behavior at small frequencies,
Phys. Rev. E {\bf 102}, 042127 (2020).

\bibitem{tor14}
E. J. Torres-Herrera and L. F. Santos,
Local quenches with global effects in interacting quantum systems,
Phys. Rev. E {\bf 89}, 062110 (2014).

\bibitem{bre20}
M. Brenes, T. LeBlond, J. Goold, and M. Rigol,
Eigenstate Thermalization in a Locally Perturbed Integrable System,
Phys. Rev. Lett. {\bf 125}, 070605 (2020).

\bibitem{san20}
L. F. Santos, F. P\'erez-Bernal, and E. J. Torres-Herrera,
Speck of chaos,
Phys. Rev. Research {\bf 2}, 043034 (2020).

\bibitem{tou15}
H. Touchette,
Equivalence and nonequivalence of ensembles: thermodynamic, macrostate, and measure levels,
J. Stat. Phys. {\bf 159}, 987 (2015).

\bibitem{bra15}
F. G. S. L. Brand\~{a}o and M. Cramer,
Equivalence of Statistical Mechanical Ensembles for Non-Critical Quantum Systems,
arXiv:1502.03263 (2015).

\bibitem{rig14}
M. Rigol,
Quantum quenches in the thermodynamic limit,
Phys. Rev. Lett. {\bf 112}, 170601 (2014).

\bibitem{rig16}
M. Rigol,
Fundamental Asymmetry in Quenches Between Integrable and Nonintegrable Systems,
Phys. Rev. Lett. {\bf 116}, 100601 (2016).

\bibitem{mal18}
K. Mallayya and M. Rigol,
Quantum Quenches and Relaxation Dynamics in the Thermodynamic Limit,
Phys. Rev. Lett. {\bf 120}, 070603 (2018).

\bibitem{dun21}
J. Dunlop, O. Cohen, and A. J. Short,
Eigenstate thermalization on average,
Phys. Rev. E {\bf 104}  024135 (2021).

\bibitem{lie72}
E. H. Lieb and D. W. Robinson,
The finite group velocity of quantum spin systems,
Commun. Math. Phys. {\bf 28}, 251 (1972).

\bibitem{has10}
M. Hastings, {\em Locality in quantum systems} in {\em Quantum Theory from Small to Large Scales: Lecture Notes of the Les Houches Summer School}, Vol. 95 (Oxford University Press, 2010).

\bibitem{bou15}
G. Bouch,
Complex-time singularity and locality estimates for quantum lattice systems,
J. Math. Phys. {\bf 56}, 123303 (2015).

\bibitem{rei18}
P. Reimann,
Dynamical typicality of isolated many-body quantum systems,
Phys. Rev. E {\bf 97}, 062129 (2018).

\bibitem{bal19}
B. N. Balz, J. Richter, J. Gemmer, R. Steinigeweg, and P. Reimann,
Dynamical typicality for initial states with a preset measurement 
statistics of several commuting observables,
Chapter 17 (p. 413-433) in {\em Thermodynamics in the Quantum Regime}, 
edited by F. Binder, L. A. Correa, C. Gogolin, J. Anders, and G. Adesso 
(Springer, Cham, 2019).

\bibitem{alb15}
V. Alba,
Eigenstate thermalization hypothesis and integrability in quantum spin chains,
Phys. Rev. B {\bf 91}, 155123 (2015).

\bibitem{yos18}
T. Yoshizawa, E. Iyoda, and T. Sagawa,
Numerical large deviation analysis of eigenstate thermalization hypothesis,
Phys. Rev. Lett {\bf 120}, 200604 (2018).



\end{thebibliography}
\end{document}